\documentclass[a4paper,11pt]{article}
\usepackage[utf8]{inputenc}
\pdfoutput=1 

\usepackage{jcappub} 
\usepackage{macros}

\usepackage{parskip}

\usepackage[dvipsnames]{xcolor}
\usepackage{graphicx}

\usepackage{caption}
\usepackage{subcaption}

\usepackage[T1]{fontenc} 

\newcommand{\erf}{\mathrm{erf}}

\newcommand{\trh}{{T}_{\mathrm{RH}}}
\newcommand{\arh}{{a}_{\mathrm{RH}}}
\newcommand{\krh}{{k}_{\mathrm{RH}}}
\newcommand{\rh}[1]{{#1}_{\mathrm{RH}}}
\newcommand{\gs}[1]{g_{*}(#1)}
\newcommand{\gss}[1]{g_{*S}(#1)}

\newcommand{\ths}{T_{\rm hs}}
\newcommand{\beqa}{\begin{eqnarray}}
\newcommand{\eeqa}{\end{eqnarray}}

\newcommand{\aeq}{a_{\rm eq}}

\newcommand{\dc}{\delta_{\rm c}}
\newcommand{\fmax}{f_{\rm max}}

\usepackage{soul}

\let\vec\bm

\newcommand{\diff}{\ensuremath{\mathrm{d}}}

\newcommand{\ar}{a_{\rm ref}}

\newcommand{\ac}{a_{\rm coll}}
\newcommand{\cs}{\langle \sigma v \rangle}
\newcommand{\nxe}{n_{X,\rm{eq}}}
\newcommand{\rcusp}{r_{\rm cusp}}
\newcommand{\rcore}{r_{\rm core}}

\usepackage[numbers]{natbib}
\usepackage{bm}
\let\vec\bm

\newcommand{\e}{\mathrm{e}}

\usepackage{amsmath}
\usepackage{multirow}

\title{\boldmath Limits on Early Matter Domination from the Isotropic Gamma-Ray Background}
\author[a,1]{Himanish Ganjoo\note{Corresponding author.}}
\author[b]{and M. Sten Delos}

\affiliation[a]{Laboratoire d’étude de l’Univers et des phénomènes eXtrêmes,
Université Paris Cité, Observatoire de Paris, Université PSL,
CNRS,
F-92190 Meudon, France}
\affiliation[b]{Carnegie Observatories, 813 Santa Barbara Street, Pasadena, CA 91101, USA}

\emailAdd{Himanish.Ganjoo@obspm.fr}

\abstract{
In cosmologies with hidden sector dark matter, the lightest hidden sector species can come to dominate the energy budget of the universe and cause an early matter-dominated era (EMDE). EMDEs amplify the matter power spectrum on small scales, leading to dense, early-forming microhalos which massively boost the dark matter annihilation signal. We use the Fermi-LAT measurement of the isotropic gamma-ray background to place limits on the parameter space of hidden sector models with EMDEs. We calculate the amplified annihilation signal by sampling the properties of prompt cusps, which reside at the centers of these microhalos and dominate the signal on account of their steep $\rho\propto r^{-3/2}$ density profiles. We also include the portions of the parameter space affected by the gravitational heating that arises from the formation and subsequent destruction of nonlinear structure during the EMDE. We are able to rule out significant portions of the parameter space, particularly at high reheat temperatures. Long EMDEs remain poorly constrained despite large structure-induced boosts to the annihilation signal.
}

\begin{document}
\raggedbottom
\maketitle
\setlength{\abovedisplayskip}{4pt}
\setlength{\belowdisplayskip}{4pt}

\section{Introduction}

Despite strong astrophysical evidence for the presence of dark matter (DM), its particle properties are unknown. Attempts to detect dark matter via direct detection experiments and collider searches have proven unsuccessful \cite{dmx1,dmx3,dmx2,partx1,partx2,partx3}, prompting renewed interest in theories in which DM resides in a ``hidden'' sector secluded from the Standard Model (SM) \cite{hid7,hid2,hid6,hid3,hid1,hidx}. Although direct detection is difficult in these models since the hidden sector is coupled very feebly to the SM, these scenarios can leave detectable signatures in the small-scale structure of the universe. 

Many hidden sector theories involve a period of early matter or cannibal domination that enhances the matter power spectrum on very small scales \cite{zhang,hid4,hid5,codm1,codm2,ae2020}. Such enhancements transform the small-scale distribution of DM by causing the early formation of small, highly dense structures. Owing to their high densities, these structures survive and populate bigger current-day halos as substructure. These microhalos can be potentially detected via their gravitational lensing signatures \cite{obs_blinov} or their distortion of pulsar timing array signals \cite{pta1,pta_sten}; they can also massively boost the potential annihilation signal from dark matter \cite{2015PhRvD..92j3505E,aew16,blanco19,sten_gr}. 

In the standard picture of cold dark matter (CDM) cosmology, bound structures form by undergoing accretion and mergers to hierarchically form larger halos from smaller ones. However, recent research has challenged this idea in cases where DM perturbations are smoothed below a certain length scale. In these cosmologies, the first bound structures form directly from the monolithic collapse of peaks in the density field. These ``prompt cusps'' have density profiles of logarithmic slope $d \ln \rho / d \ln r = -3/2$ \cite{cusp5,cusp6,cusp7,cusp10,cusp1,cusp3,cusp9,cusp4,cusp8,cusp2,sten_survival}, which are markedly different from the Navarro-Frenk-White (NFW) \cite{nfw96} profiles seen in CDM halos formed in $N$-body simulations. Despite containing $\mathcal{O}(1\%)$ of the dark matter, these cusps radically alter the DM annihilation signal, which they dominate on account of their steep density profiles \cite{sten_white_cusp}.   

In this paper, we calculate the DM annihilation signal arising from cusps in cosmologies with an early matter-dominated era (EMDE). Subhorizon matter perturbations grow linearly with scale factor during an EMDE, enhancing the matter power spectrum on scales which enter the horizon during or before the EMDE \cite{emde2,emde1,emde3,ae15}. 
The matter power spectrum can develop a small-scale cut-off owing to the thermal motion of the EMDE-causing particle \cite{hg23} or the dark matter, or due to the free-streaming of dark matter from the destruction of microhalos that form during the EMDE \cite{heating}. In all of these cases, the smoothing of matter perturbations below the cut-off scale will lead to the formation of prompt cusps. Since the EMDE-enhanced power spectrum exhibits a $k^4$ rise on small scales, the cusps forming in such cases collapse much earlier and are thus much denser than those in standard $\Lambda$CDM cosmologies, leading to much higher boosts to the DM annihilation signal than in scenarios without EMDEs. 

Dark matter annihilation or decay could contribute to the isotropic gamma-ray background (IGRB) signal detected by the Fermi-LAT telescope \cite{fermilat}, making it an avenue for constraining the particle properties of DM. We use a modified version of the \textsc{cusp-encounters} code \cite{cuspe} to sample prompt cusps in a large variety of EMDE cosmologies, calculating the annihilation signal from these samples in each case. The morphology of the DM annihilation signal from cusps is similar to that arising from decaying dark matter. We leverage this connection to constrain the dark matter annihilation cross-section for each EMDE case using the constraints on the decaying dark matter lifetime from the IGRB signal \cite{decaydm}. 

We test a broad range of EMDE parameters.
By combining limits from the IGRB with the unitarity limit for point-like DM, we are able to place strong constraints on EMDEs that are either short or driven by very heavy particles. These scenarios tend to be associated with high reheat temperatures.
However, significant regions of the parameter space remain allowed by the unitarity bound and the IGRB, particularly in the regime where gravitational heating is important, even though they yield cusps that are orders of magnitude denser than
standard $\Lambda$CDM cusps. A similar analysis was done in Ref.~\cite{blanco19}, but our work improves upon that by including the exact small-scale cut-off due to the pressure of the particle causing the EMDE and the effects of gravitational heating. Moreover, our analysis calculates the substructure boost to the DM annihilation signal using the machinery of prompt cusps. Reference~\cite{sten_gr} also derived their constraints similarly, but with arbitrary cut-offs on the matter power spectrum. In this work, we connect observational constraints directly to the properties of the EMDE and the hidden sector particle that causes it. 

This paper is organized as follows. In Section \ref{emde_back} we describe our EMDE model including the evolution of the homogeneous background, the EMDE-enhanced power spectrum, and the concept of gravitational heating. Section \ref{cusps} describes prompt cusps and how their properties relate to the peaks in the matter density field. In Section \ref{code}, we describe the \textsc{cusp-encounters} code and the changes implemented in it to include EMDE cosmologies and gravitational heating; we also compare the properties of cusps forming in cosmologies with and without EMDEs and calculate the effects of tidal stripping and stellar encounters on the EMDE cusps. We present our results and constraints in Section \ref{results} and summarize our work in Section \ref{summary}. We work with natural units throughout, in which $c = \hbar = k_B = 1$.

\section{Early Matter Domination}
\label{emde_back}

According to current constraints, the universe was radiation dominated from at least 
a temperature of a few MeV \cite{2015PhRvD..92l3534D,2019JCAP...12..012H}, corresponding to the time of neutrino decoupling shortly before the onset of Big Bang nucleosynthesis.
However, earlier times are poorly understood \cite{fts-rev,Batell:2024dsi} and could include an epoch when a matter-like species in the hidden sector dominates the universe. In this section, we outline our model of a universe with an early matter-dominated era (EMDE) which is caused by the lightest massive hidden sector particles dominating the universe. We describe the evolution of the cosmological background, perturbation evolution and the concept of gravitational heating.
We parametrize our model as in Ref.~\cite{hg23}. 

\subsection{Evolution of the Cosmological Background}

In our model, the universe is populated by three species: the Standard Model radiation in the visible sector at temperature $T$, along with the dark matter (DM or $X$) and the $Y$ particles in the hidden sector, which has its own temperature $T_{\rm hs}$. The $Y$ particles are the lightest hidden sector species and are initially relativistic, becoming nonrelativistic as $T_{\rm hs}$ decreases and approaches their mass. After the $Y$ particles become nonrelativistic, their energy density drops slower than that of SM radiation and they come to dominate the universe to cause an EMDE. These $Y$ particles are unstable; they decay into SM radiation with a rate $\Gamma$, acting as a mediator for dark matter annihilation, which proceeds as $XX \rightarrow YY$. When the Hubble rate becomes comparable to $\Gamma$, the $Y$ particle decays become significant and their number density decreases rapidly, restoring radiation domination. After this process, which we call reheating, the thermal history of the universe follows that of a standard $\Lambda$CDM cosmology. 

This background model is completely specified by three parameters. The mass of the $Y$ particle, $m_Y$, determines when the $Y$ particles become nonrelativistic. Secondly, we have $\eta$, which is the ratio of the SM radiation to the $Y$ particle energy densities when the $Y$ particles are initially relativistic.\footnote{$\eta$ is the ratio of SM to $Y$ energy densities after the dark matter, $X$, freezes out. At temperatures above the mass $m_X$ of the dark matter, $\rho_{\rm SM} / \rho_Y=(1 + g_X / g_Y)^{4/3}\eta$ instead, where $g_X$ and $g_Y$ are the numbers of degrees of freedom of the $X$ and $Y$ particles. This is because the $X$ annihilations have not yet heated the $Y$ particles.} This parameter sets which species dominates the energy budget of the universe before the EMDE. Finally, we define $\trh$ as the temperature of a SM-dominated universe when its Hubble rate is equal to $\Gamma$, i.e. \begin{equation}\label{gamma}
    \Gamma \equiv \sqrt{\frac{8 \pi G}{3} \frac{\pi^2}{30} \gs{\rh{T}} \rh{T}^4},
\end{equation} where $g_{*}$ is the effective number of degrees of freedom contributing to the SM radiation energy density. The parameters $\eta$ and $m_Y$ together determine when the EMDE begins,
whereas $\trh$ controls when the EMDE ends.\footnote{The EMDE cosmology and the power spectrum also depend on the statistics of the $Y$ particles. The results in this paper are for scalar boson $Y$ particles, but can be easily extended to general $Y$ using the results described in Ref.~\cite{hg23}. }

Reference \cite{hg23} also defined the scale factor $\arh$ associated with reheating, connecting it to the scale factor today. This connection was established by noting that the $Y$ particle decay into the visible sector is negligible after about $5 \arh$ and conserving the visible sector entropy from $5 \arh$ to today. However, this choice of the connecting scale factor is arbitrary; it introduces a $\gss{0.204\trh}$ factor in the expression for $\arh$, where $g_{*S}$ is the effective number of degrees of freedom contributing to the SM radiation entropy. To avoid the numerical artifacts associated with this arbitrary choice, we use a modified definition, given by \begin{equation}\label{arha0}
    \frac{\rh{a}}{a_0} = \frac{1}{1.02} \left[ \frac{\gss{T_0}}{\gss{\rh{T}}} \right]^{\frac{1}{3}} \left[ \frac{T_0}{\rh{T}} \right] ,
\end{equation} where $a_0$ and $T_0$ are the scale factor and the radiation temperature today, respectively. Note that both $\rh{T}$ and $\rh{a}$ are defined quantities and $T(\rh{a}) \neq \rh{T}$. Away from SM phase transitions, Eq.~(\ref{arha0}) is the same as the definition in Ref.~\cite{hg23}. However, if there is a phase transition, this new definition ensures that $\arh$ and $\trh$ are evaluated consistently either both before the transition or both after it. We must note that results close to phase transitions (like the QCD phase transition at $T \simeq 170$ MeV) are approximate. 

\begin{figure}[h!]
\centering
\includegraphics[scale=1.3]{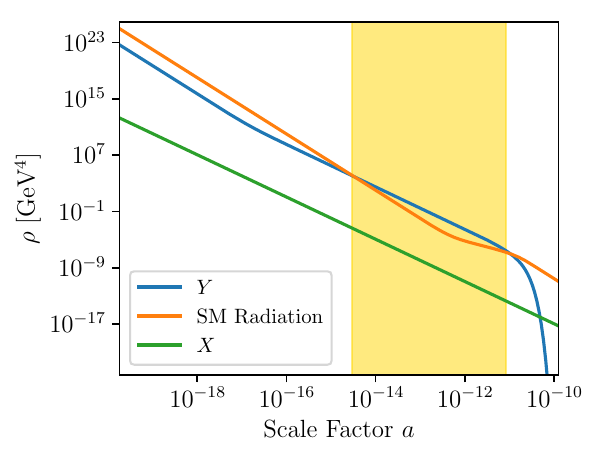}
\caption{The background evolution of the energy densities of the $Y$ particles, SM radiation and dark matter ($X$) as a function of scale factor. The background cosmology is specified by $m_Y = 2$ TeV, $\eta = 200$, $\trh = 20$ MeV. The yellow shaded region shows the EMDE.}
\label{fig:back}
\end{figure}

In Figure \ref{fig:back}, we show the evolution of the energy densities of these three fluids versus scale factor $a$ for an example EMDE case with $m_Y = 2$ TeV, $\eta = 200$, $\trh = 20$ MeV. The blue curve shows $\rho_Y$, which is proportional to $a^{-4}$ initially but transitions to $a^{-3}$ behavior at a scale factor of about $10^{-17}$. The yellow shaded region shows the EMDE, which begins at $a \approx 10^{-15}$ when $\rho_Y$ exceeds $\rho_{\rm SM}$. Radiation domination is restored at $a \approx 10^{-11}$ as the $Y$ particle population decays away rapidly. The dark matter has frozen out and $\rho_X \propto a^{-3}$. 

For more details on the EMDE model parameters, the background model and the methods used to solve for the time evolution of thermal history of the hidden sector, we refer the interested reader to Ref. \cite{hg23}.

\subsection{Perturbation Evolution and Power Spectrum}

\begin{figure}[h!]
\centering
\includegraphics[scale=1.4]{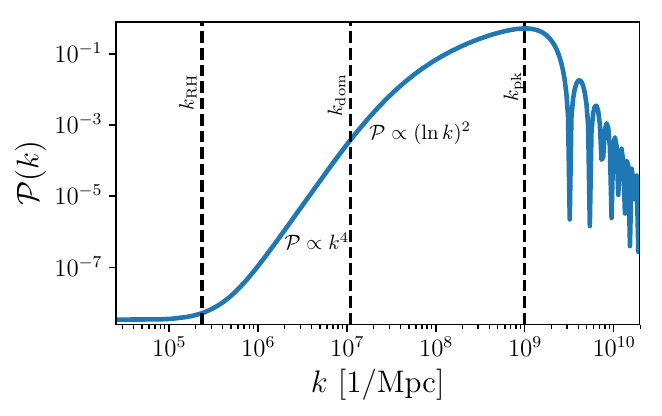}
\caption{The linear-theory dark matter power spectrum at the end of the EMDE for a case given by $m_Y = 2$ TeV, $\eta = 500$, $\trh = 20$ MeV. The wavenumber $\krh$ indicates the mode which enters the horizon at the end of the EMDE, while $k_{\rm dom}$ is the horizon wavenumber at the start of the EMDE. }
\label{fig:ps}
\end{figure}

A period of matter domination causes subhorizon matter perturbations to grow linearly with scale factor, in contrast to the logarithmic growth seen during radiation domination \cite{emde2,emde1,emde3,ae15}. Consequently, density perturbations for modes which enter the horizon before or during the EMDE are boosted compared to cosmologies without an EMDE. To illustrate this, Figure \ref{fig:ps} shows the dimensionless power spectrum of dark matter perturbations $\mathcal{P}(k) \equiv k^3 P(k) / (2\pi^2)$ for an EMDE cosmology with $m_Y = 2$ TeV, $\eta = 500$ and $\trh = 20$ MeV, calculated at the end of the EMDE using linear theory. 

The modes with $k_{\rm dom} < k < \krh$ enter the horizon during the EMDE and grow linearly with scale factor, which translates to a $k^4$ rise in the power spectrum.
Here $k_{\rm dom}$ and $\krh$ are the wavenumbers entering the horizon at the beginning and end of the EMDE, respectively.
The modes with $k>k_{\rm dom}$ enter during the radiation epoch before the EMDE, growing logarithmically with scale factor before the EMDE and linearly during the EMDE. This growth pattern makes the power spectrum proportional to $(\ln k)^2$ for these modes. Modes with $k>k_{\rm pk}$,
where $k_{\rm pk}$ is the location of the maximum in the power spectrum,
enter the horizon while the $Y$ particles have significant pressure due to being relativistic. The growth of $Y$ perturbations is inhibited until the $Y$ particles become cold, after which their evolution mirrors that of a cold dark matter perturbation. We assume that the $Y$ particles do not undergo number-changing interactions (e.g. cannibals \cite{cannibal_big}), which can lead to a different perturbation evolution.

This suppression of $Y$ perturbations manifests as a fall-off with decaying oscillations in the dark matter power spectrum. During the EMDE, the $Y$ particles dominate the universe and cluster to form gravitational wells. The DM particles fall into these wells, clustering with the $Y$ particles. As a result, the DM density perturbations track the $Y$ density perturbations. The power spectra of $Y$ and DM perturbations are thus nearly equal for all $k > \krh$ by the end of the EMDE. Since the $Y$ particles are much lighter than the DM in hidden-sector theories \cite{hid7}, the dark matter velocities are much lower than the $Y$ particle velocities. The DM free-streaming length is therefore very small compared to the Jeans length of the $Y$ particles, so the latter sets the cut-off on the power spectrum.

\begin{figure}[h!]
\centering
\includegraphics[scale=1.0]{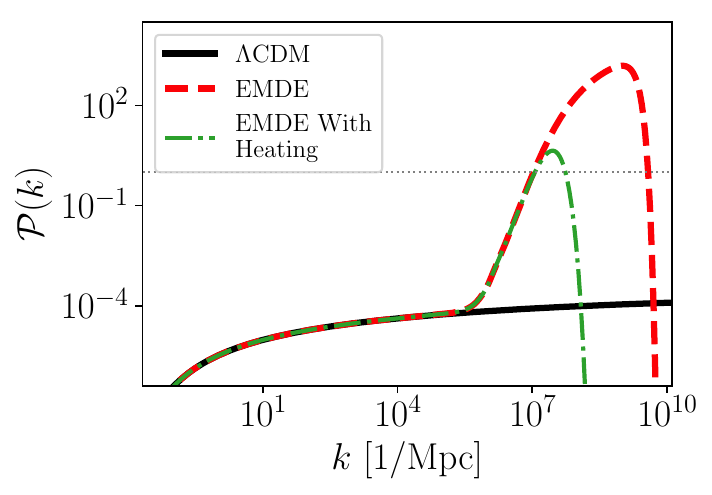}
\caption{Dashed curve: Linear theory EMDE power spectrum at the time of matter-radiation equality, computed using the transfer functions from Ref. \cite{hg23} for a case given by $m_Y = 2$ TeV, $\eta = 500$, $\trh = 20$ MeV. Solid curve: no-EMDE $\Lambda$CDM power spectrum at the same time for comparison. Dash-dotted curve: The EMDE power spectrum with the gravitational heating cut-off imposed. The dotted horizontal line marks $\mathcal{P} = 1$; $\mathcal{P} \gtrsim 1$ implies the formation of nonlinear structures on those scales.}
\label{fig:psemde}
\end{figure}

Reference \cite{hg23} provided accurate fitting forms for the DM power spectrum after a $Y$-driven EMDE; we will use these functions for the rest of this paper. Figure \ref{fig:psemde} shows the EMDE power spectrum at matter-radiation equality (dashed curve), computed using the transfer functions for $m_Y = 2$ TeV, $\eta = 500$ and $\trh = 20$ MeV. The solid line shows the $\Lambda$CDM power spectrum in a cosmology without an EMDE at the same time for comparison. The transfer functions are designed to accurately model the power spectrum close to where it is maximized, ignoring the oscillating fall-off at larger $k$, which is relatively unimportant for structure formation owing to its smaller amplitude. 

\subsection{Gravitational Heating}

In cases where the EMDE is long enough, matter perturbations can grow to become nonlinear before the EMDE ends, causing the formation of collapsed structures \cite{blanco19}. For instance, for the case shown in Figure \ref{fig:ps}, Press-Schechter theory \cite{press-schechter} predicts that 22\% of matter will be bound in halos at the end of the EMDE. These structures are composed mostly of the $Y$ particles, which decay at the end of the EMDE, evaporating the bound objects as a result. The DM particles which have fallen into these structures following the $Y$ particles experience a sudden loss of gravitational potential and are shot out at their virial velocities. This process of structure formation and dissolution boosts the DM particles' speeds and randomizes their directions, effectively heating up the dark matter. 

This gravitational heating causes the dark matter to free-stream shortly after the EMDE ends, imposing a time-evolving cut-off on the power spectrum after the EMDE. Although it results from the complex process of nonlinear structures forming and evaporating, we showed in Ref. \cite{heating} that the cut-off scale can be calculated from the linear power spectrum resulting from the EMDE, and we supplied two kinds of transfer functions to encode this cut-off. 

The dash-dotted curve in Fig. \ref{fig:psemde} shows the EMDE power spectrum with the cut-off from gravitational heating.\footnote{Figure~\ref{fig:psemde} still shows the linear-theory power spectrum, in the sense that although the power spectrum incorporates the free-streaming cut-off from gravitational heating -- a consequence of nonlinear evolution -- the power spectrum is otherwise evolved in accordance with linear theory.} Here and for the remainder of this work, we use the exponential cut-off (instead of the power law cut-off) because it produces a sharp fall-off, which is amenable to our sampling procedure (this is explained further in Section \ref{code}). This power spectrum reaches a value greater than 1 at the epoch of matter-radiation equality ($z \approx 3400$), which indicates the formation of bound structures at this early time even after the gravitational heating erases much of the EMDE-induced bump in the power spectrum. These early-forming structures are much denser than later-forming halos, because the central density of a bound object scales as $a_f^{-3}$, where $a_f$ is the formation scale factor. On account of their high central densities, these early-forming structures will massively boost the DM annihilation signal.

\section{Prompt Cusps}
\label{cusps}

Recent studies have shown that the first dark matter structures, forming from the collapse of smooth density peaks, have density profiles $\rho \propto r^{-3/2}$ \cite{cusp5,cusp6,cusp7,cusp10,cusp1,cusp3,cusp9,cusp4,cusp8,cusp2,sten_survival}, which are much steeper than the Navarro-Frenk-White (NFW) profiles that provide good fits to the profiles of halos seen in $N$-body simulations of much larger scales. Owing to hierarchical mass accretion and mergers, larger structures with the NFW profiles assemble around these prompt cusps, but the cusp profile still persists in the inner regions \cite{sten_survival}. As a result, these cusps populate the substructure of current-day halos and greatly enhance the annihilation signal from them. In this section, we describe how to sample a population of cusps and describe their density profiles, thus setting up the machinery to calculate their annihilation signal. 

Since the cusp forms almost immediately after the collapse of a density peak, cusp properties can be easily inferred from the peak properties. Reference \cite{sten_survival} found that these cusps have a profile $\rho = Ar^{-3/2}$, with the prefactor well approximated by \begin{equation} \label{Acusp}
    A = 24 \rho_0 \ac^{-3/2} R^{3/2},
\end{equation} where $\rho_0$ is the current dark matter density, $\ac$ is the scale factor of the peak's collapse and $R = \sqrt{\delta / |\nabla^2 \delta|}$ with $\delta$ being the amplitude of the linear overdensity of the peak. Moreover, Ref. \cite{sten_survival} found that the cusp profile extends out to an approximate radius \begin{equation} \label{rcusp}
    \rcusp = 0.11 \ac R.
\end{equation}
These results are valid for collapse during matter domination, but we expect them to also hold for peaks that collapse shortly before matter domination, since the collapsing density peak would become locally matter dominated.

The collapse scale factor $\ac$ for a peak that has amplitude $\delta(a)$ at some scale factor $a$ is given by the relation \begin{equation} \label{peak_col}
    \dc (\ac;e,p) = \frac{D(\ac)}{D(a)} \delta(a),
\end{equation} where $D(a)$ is the linear growth function, $\dc$ is the linear density threshold for collapse and $e$ and $p$ are the ellipticity and prolateness of the tidal field at the location of the peak. In Section~\ref{code}, we discuss how $\ac$, $e$, and $p$ are factored into the collapse criterion in our calculations.

\subsection{The Cusp Distribution}
\label{cuspdis}

The distribution of cusp properties can be calculated from the linear power spectrum of matter density perturbations $P(k)$ owing to the peak-cusp connection given by Eqs.~(\ref{Acusp}) and (\ref{rcusp}). 

The distributions of $R$ and $\ac$ can be found from the distributions of $\delta$, $\nabla^2 \delta$, $e$, and $p$, which can be computed from the power spectrum by using the statistics of density peaks \cite{bbks}. If $\sigma_j^2 \equiv \int \mathcal{P}(k) k^{2j} \mathrm{d} \ln k$, $\gamma = \sigma_1^2 / (\sigma_0 \sigma_2)$ and $R_{*} \equiv \sqrt{3} \sigma_1 / \sigma_2$, the differential number density of peaks is given by \begin{equation}\label{nux}
\frac{\mathrm{d}^2n_\mathrm{peaks}}{\mathrm{d}\nu\mathrm{d}x}
=
\frac{\mathrm e^{-\nu^2/2}}{(2\pi)^2R_*^3}f(x)\frac{\exp\!\left[-\frac{1}{2}(x-\gamma\nu)^2/(1-\gamma^2)\right]}{[2\pi(1-\gamma^2)]^{1/2}},
\end{equation} where the parameters $\nu \equiv \delta / \sigma_0$ and $x \equiv -\nabla^2 \delta / \sigma_2$ are the scaled peak height and curvature, respectively, and \begin{equation}
    f(x)
\equiv
\frac{x^3\!-\!3x}{2}\left[\erf\!\left(\sqrt{\frac{5}{2}}x\right)\!+\!\erf\!\left(\sqrt{\frac{5}{8}}x\right)\right]
\!+\!
\sqrt{\frac{2}{5\pi}}\left[\left(\frac{31}{4}x^2\!+\!\frac{8}{5}\right)\mathrm e^{-\frac{5}{8}x^2}\!+\!\left(\frac{x^2}{2}\!-\!\frac{8}{5}\right)\mathrm e^{-\frac{5}{2}x^2}\right]. 
\end{equation} Finally, the ellipticity and prolateness distribution for peak height $\nu$ is approximately \cite{sheth_mo_tormen} \begin{equation}
    \label{ep}
f(e,p\mid\nu) = \frac{1125}{\sqrt{10\pi}}e(e^2-p^2)\nu^5\exp\left[-\frac{5}{2}\nu^2(3e^2+p^2)\right].
\end{equation}

\subsection{The Inner Core}
\label{core}

The $r^{-3/2}$ profile does not extend down to the center of the cusp. Liouville's theorem dictates that the maximum phase space density $\fmax$ of the dark matter sheet cannot grow in time \cite{gunn_tremaine}. Consequently, $\fmax$ set before the formation of bound structures cannot be exceeded in any non-linear structure later. This limit implies the presence of a central core with a constant density $\rho_{\rm core} = A \rcore ^{-3/2}$. Assuming that the phase-space limit is saturated in the core (as suggested by the simulations of Ref.~\cite{2012MNRAS.424.1105M}), the expression for the core radius is \cite{sten_survival} \begin{equation}
    \rho_{\rm core} \rcore^6 = 3 \times 10^{-5} G^{-3} \fmax^{-2},
\end{equation} where $G$ is the gravitational constant.\footnote{Annihilation would separately impose a maximum density $\rho_{\rm max}\sim m_X/(\cs t)$, where $\cs$ is the dark matter annihilation cross section and $t$ is the age of the Universe. However, for the scenarios that we consider, this maximum density vastly exceeds the core densities $\rho_{\rm core}$. For example, for an EMDE scenario with $m_Y = 2$ TeV, $\eta = 10^3$, and $\trh = 10$ MeV, the $\cs$ and minimum $m_X$ that we obtain in Fig.~\ref{fig:svp_heat} is associated with $\rho_{\rm max}\sim 10^{15}$~M$_\odot$pc$^{-3}$, which is far higher than the $\rho_{\rm core}\sim 10^7$~M$_\odot$pc$^{-3}$ that arise in this scenario.}

The value of $\fmax$ is set by the width of the phase-space sheet of the dark matter. If dark matter experiences gravitational heating, this width depends only on the growth of structure during the EMDE.\footnote{Although the fine-grained phase-space density is conserved in principle throughout structure formation, phase mixing in the (nonlinear) multi-streaming regime suppresses the coarse-grained phase-space density both during and after the EMDE. Some memory of the original $\fmax$ could be preserved in matter that did not fully phase mix, but we conservatively neglect that consideration, which would only apply to a fraction of the mass, and we assume that the heating-induced streaming velocities impose a new effective $\fmax$.} However, without heating, the width is controlled by dark matter freeze-out. If dark matter freezes out when nonrelativistic, it has a momentum distribution $f_p (\textbf{p}) = (2 \pi p_0^2)^{-3/2} \exp [-\textbf{p}^2 / (2p_0^2)]$, which is maximal at $\textbf{p} = 0$ and has a standard deviation $p_0$. The maximum value of $f_p$ is $(2 \pi p_0^2)^{-3/2}$. With freeze-out at scale factor $a_f$ with hidden sector temperature $T_f$, we have $p_0 = p_f (a_f/a)$, where $p_f = (T_f m_X)^{1/2}$. This implies that the maximum of the momentum distribution is $(2 \pi)^{-3/2} a^3 / (p_f a_f)^3$. If $\rho_X (a) = \rho_m a^{-3}$ (where $\rho_m$ is the dark matter density today), the maximum position-momentum phase space density $\fmax = (2 \pi)^{-3/2} \rho_m (p_f a_f)^{-3}$. The code we use in this paper (described in Sec.~\ref{code}) works with $\fmax$ in position-velocity phase space, where it is \begin{equation}
    \fmax = (2 \pi)^{-3/2} \rho_m (p_f a_f / m_X)^{-3}.
\end{equation} In Appendix \ref{app:dmfo}, we calculate the scale factor and temperature of dark matter at freeze-out if it freezes out in a hidden sector with an EMDE. 

\subsection{Annihilation In Cusps}
\label{cusps_ann}

Assuming a velocity-independent DM annihilation cross-section, the annihilation signal from each cusp is proportional to \begin{equation} \label{jinteg}
    J = \int \rho^2 \textrm{d} V,
\end{equation} where $\rho$ is its density profile. Modeling the density profile as \begin{equation}
    \rho(r) = \begin{cases}
    A \rcore^{-3/2} & r < \rcore \\
    A r^{-3/2} & \rcore \leq r < \rcusp,
    \end{cases}
\end{equation} the right side of Eq.~(\ref{jinteg}) integrated from $r = 0$ to $r = \rcusp$ gives $J = 4 \pi A^2 [0.333 + \ln (\rcusp / \rcore)]$. The \textsc{cusp-encounters} code we employ in this work uses $J = 4 \pi A^2 [0.531 + \ln (\rcusp / \rcore)]$, which results from a smoother and mass-conserving model of the transition of the core-to-cusp profile.  

In some cases, the upper limit on the DM phase-space density, set by the thermal motion of the DM at freeze-out or the velocity dispersion due to gravitational heating, leads to core radii that are greater than cusp radii if the $Ar^{-3/2}$ density profile is used to obtain $\rcore$. Physically, this means that the maximum phase-space density is too low for the cusp to form. However, the collapsed structure surrounding a core still develops. In these cases, we make a conservative estimate of the surrounding halo. We assume a density profile that transitions from $\rho=Ar^{-3/2}$ at small $r$ to $\rho\propto r^{-3}$ at large $r$ and impose the phase-space density condition on this new profile to evaluate $\rcore$. We replace the $J$ from these cusps with the $J$ evaluated using the new multiple-power-law profile; the details are presented in Appendix~\ref{app:invcore} along with a demonstration that it is a conservative choice.
 
\section{Cusp Sampling Code: Modifications and Results}
\label{code}

To sample the properties of cusps from EMDE-enhanced power spectra, we use a modified version of the \textsc{cusp-encounters} code \cite{cuspe}. This code uses CLASS \cite{classcode} to generate a linear power spectrum at a reference redshift $z_{\rm ref}$ up to the wavenumber $k = 10^4$ $h/$Mpc (where $h\simeq 0.68$ is the Hubble parameter). For smaller scales, it uses the analytical expressions for the power spectrum given by Hu \& Sugiyama \cite{hu96} to model the non-clustering of baryons. The power spectra in the two regimes are smoothly matched at $k = 10^4$ $h/$Mpc. The peak properties are sampled using this power spectrum from the distributions described in Sec.~\ref{cuspdis} and the cusp properties are calculated using Eqs.~(\ref{Acusp}) and (\ref{rcusp}). We make several modifications to this code to calculate the properties of cusps in EMDE cosmologies: \begin{enumerate}
    \item \textbf{EMDE Transfer Functions}: The original code features cut-off functions for weakly interacting massive particle (WIMP) models and warm dark matter models. We added the EMDE transfer functions given in Ref.~\cite{hg23}. These functions accurately model the $k^4$ rise in the power spectrum for modes which entered the horizon during the EMDE, the $(\ln k)^2$ feature in the power spectrum for modes which entered the horizon before the EMDE, and the small-scale cut-off arising from the relativistic pressure of the $Y$ particles. 
    
    \item \textbf{Heating Transfer Functions}: We implemented the gravitational heating-based power spectrum cut-off in Ref.~\cite{heating} for cases in which enough matter ($> 1$\%) is bound in halos when the EMDE ends, which is the regime in which this prescription is valid.
    Moreover, we chose the exponential cut-off instead of the $k^{-5}$ function in Ref.~\cite{heating} because the sampling procedure works best for sharp cut-offs. Both the power law and exponential cut-off forms match the simulation power spectra in Ref.~\cite{heating}, but the latter has further motivation because it is derived from the velocity distribution of the DM particles in the simulations. Additionally, the peak sampling procedure can fail for shallow cut-offs, since the assumption of the peaks collapsing in isolation breaks down as the spacing between peaks becomes much smaller than the peak size.
    
    \item \textbf{Collapse Condition}: To accommodate EMDE cosmologies and peak collapse during radiation domination, we used a time-dependent linear threshold $\dc$ for spherical collapse which differs from the standard value of $1.686$ valid for collapse during matter domination;
    we implemented this using the fitting function derived in Ref.~\cite{blanco19}. Furthermore, the growth function $D(a)$ was taken from Ref. \cite{heating} and accounts for the growth of perturbations in mixed matter-radiation domination for modes that entered the horizon during or before the EMDE. This $D(a)$ also accounts for the non-clustering of baryons on small scales (e.g.~\cite{hu96,2006PhRvD..74f3509B}). 
    
    We modified the condition for the peak collapse in the code to include the time-dependent linear density threshold and the modified growth function. Since the peaks are sampled from a power spectrum calculated at a reference scale factor $a_{\rm ref} = 1/31.6$, the collapse criterion \begin{equation} \label{collcond}
         \delta_c(\ac) f_{\rm ec} (e,p) = \frac{D(\ac)}{D(\ar)} \delta (\ar)
    \end{equation} is solved numerically to find $\ac$ for each density peak. In Eq.~(\ref{collcond}), $f_{\rm ec} (e,p)$ is the ellipsoidal collapse correction factor \cite{sheth_mo_tormen}, well-approximated by \begin{equation} \label{ecoll}
    f_{\rm ec} (e,p) = 1 + 0.47 [5(e^2 - p|p|)f_{\rm ec}^2 (e,p)]^{0.615},
    \end{equation} where $e$ and $p$ are the ellipticity and prolateness of the peak. Equation~(\ref{ecoll}) has no solutions for $e^2 - p|p| > 0.26$; we assume that the peaks that match this condition do not collapse into cusps. 
    
    This method ignores the logarithmic time-variation of the gravitational heating cut-off before the epoch of matter-radiation equality, which can slightly shift the location of the maximum in the power spectrum at these times.
    Additionally, we neglect how the ellipsoidal collapse correction $f_{\rm ec}$ varies prior to matter-radiation equality (e.g.~\cite{2023MNRAS.520.4370D}).
    
    \item \textbf{Core Radii Due to Freeze-Out}: We included the calculation of the freeze-out of dark matter in a hidden sector with an EMDE (described in Appendix \ref{app:dmfo}) to set the core radii. For each EMDE case, the freeze-out condition given by Eq.~(\ref{focond}) is solved to find the hidden sector temperature at freeze-out, which is then used to determine the scale factor of freeze-out along with the width of the dark matter momentum distribution at this time. 
    
    \item \textbf{Core Radii Due to Heating}: In cases where the gravitational heating cut-off is imposed, the width of the momentum distribution depends on the velocity dispersion caused by structure formation during the EMDE. We used the velocity distribution function derived in Ref.~\cite{heating} as a fit to simulation results, \begin{equation}
        f(\textbf{v}) = 3.98 \times 10^{-11} \sigma^{-3} [e^{[|\textbf{v}|/(14.9 \sigma)]^{1.94}} - 0.953]^{-7.06},
    \end{equation}where $\sigma$ is the linear velocity dispersion. This function peaks at $\textbf{v} = 0$ with the value $0.094 \sigma^{-3}$. If the EMDE ends at $\arh$, $\sigma = \sigma_{\rm RH} \arh / a$, where $\sigma_{\rm RH}$ is the truncated velocity dispersion at the end of the EMDE (calculated as in Ref.~\cite{heating}). Consequently, in such cases $\fmax = 0.094 \rho_m (\sigma_{\rm RH} a_{\rm RH} )^{-3}$, following the calculations at the end of Section~\ref{core}. 
    
\end{enumerate}

\begin{figure}[h!]
\centering
\includegraphics[width=0.95\textwidth]{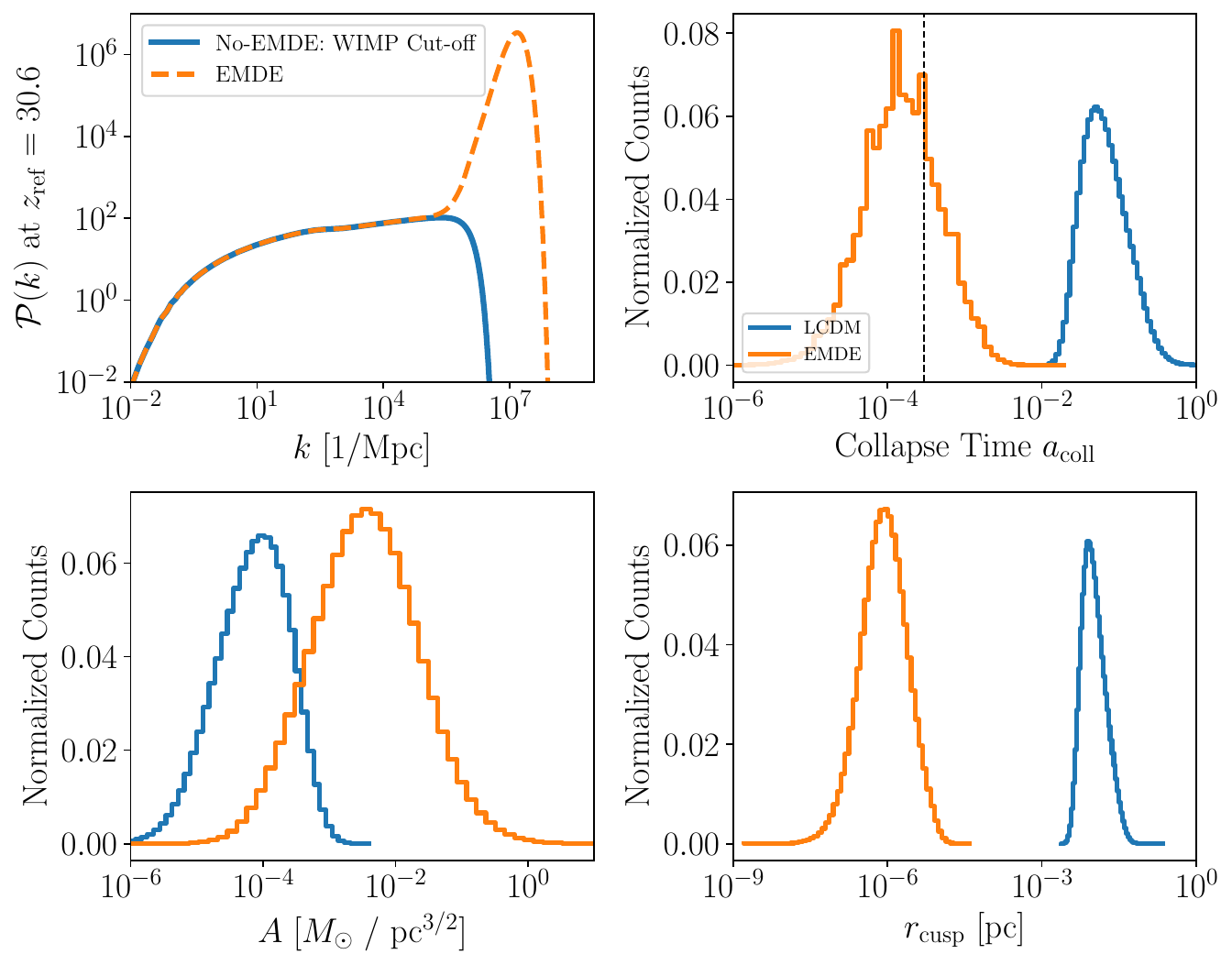}
\caption{
\textbf{Top left}:
Linear-theory dark matter power spectra for a standard $\Lambda$CDM scenario (blue) and an EMDE cosmology (dashed orange). For the $\Lambda$CDM scenario, the small-scale cut-off is due to the thermal motion of a 100 GeV WIMP. For the EMDE, we set $m_Y = 4.5$ TeV, $\eta = 1000$, and $\trh = 15.2$ MeV, and we account for gravitational heating.
\textbf{Top right}: Histograms of the collapse scale factors for sampled peaks from the $\Lambda$CDM (blue) and EMDE (orange) power spectra. Even after accounting for gravitational heating, many of the peaks are predicted to collapse even before matter-radiation equality (vertical dashed line).
\textbf{Bottom left}: Histograms of the prompt cusp density coefficients $A = \rho r^{3/2}$ for the two cases. \textbf{Bottom right}: Histograms of the cusp radii for the two cases. }
\label{fig:histcomp}
\end{figure}

We compared the properties of the cusps resulting from an EMDE cosmology with gravitational heating to those of cusps in a EMDE-less $\Lambda$CDM cosmology by sampling $N = 10^7$ density peaks in each case. For the no-EMDE case, we used the standard version of the code with the small-scale cut-off due to a WIMP of mass 100 GeV which decoupled from the thermal bath at a temperature of 30 MeV, as in Ref.~\cite{cuspe}. For the EMDE case, we use our modified code with $m_Y = 4.5$ TeV, $\eta = 1000$ and $\trh = 15.2$ MeV. The top left panel of Figure~\ref{fig:histcomp} shows the no-EMDE and EMDE linear dark matter power spectra used for the sampling runs at the reference redshift $z_{\rm ref} = 30.6$. A significant portion of the small-scale enhancement due to the EMDE is preserved even after the free-streaming cut-off due to gravitational heating is imposed. As a result, the maximum of the EMDE power spectrum is much higher than that of the no-EMDE power spectrum.

The impact of this difference can be seen in the properties of cusps in the two scenarios. The other panels of Figure~\ref{fig:histcomp} show the distributions of various cusp properties as normalized counts, comparing cusps from the two cases. The top right panel shows the distribution of collapse scale factor $\ac$ for cusps in the two cases. Since the maximum of the EMDE power spectrum is much higher than that of the no-EMDE power spectrum, the EMDE cusps collapse much earlier than the no-EMDE cusps, with a majority of them forming before the time of matter-radiation equality, which is shown by the vertical dashed line. This early collapse leads to much denser cusps in the EMDE case. The bottom right panel of Figure~\ref{fig:histcomp} shows the distributions of $\rcusp$: the EMDE cusps are much smaller than the no-EMDE counterparts because the maximum of the EMDE power spectrum is at a much smaller scale compared to the cut-off scale imposed on the $\Lambda$CDM power spectrum, and because the EMDE cusps collapse much earlier when the peaks (of set comoving size) are smaller. Finally, the bottom left panel shows the distributions of $A$, the pre-factor for the $Ar^{-3/2}$ profile. The value of $A$ increases with both the size and density of the cusps. Although the EMDE cusps are much smaller, their densities exceed the no-EMDE cusp densities by a bigger factor. As a result, the mean value of $A$ is 0.0001 $M_{\odot} / \textrm{pc}^{3/2}$ for the no-EMDE cusps and 0.02 $M_{\odot} / \textrm{pc}^{3/2}$ for the EMDE cusps. 

\begin{figure}[h!]
\centering
\includegraphics[width=0.6\textwidth]{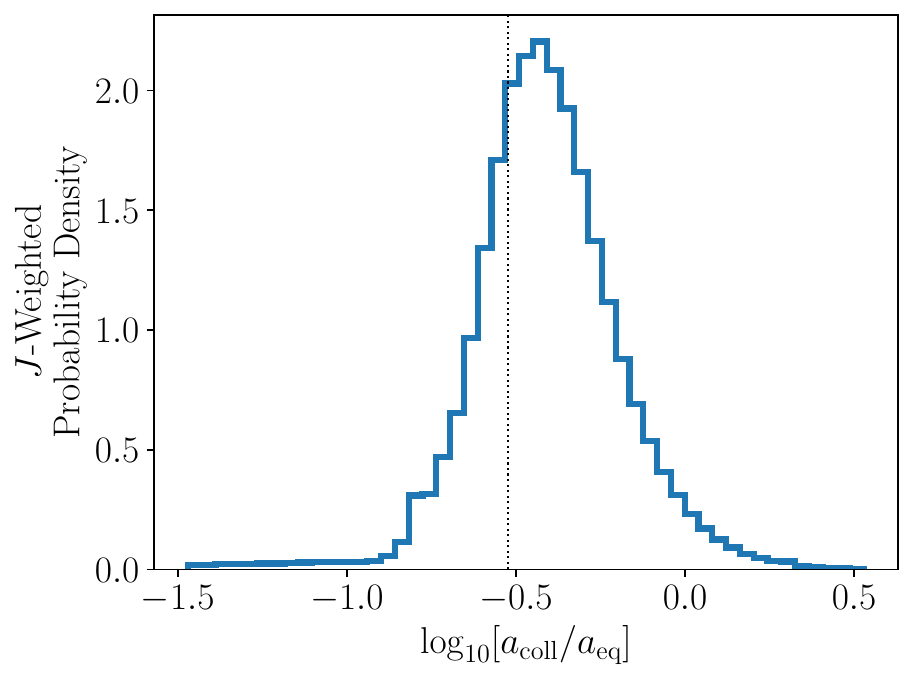}
\caption{Histogram of the collapse times of sampled peaks, weighted by the $J$-factor of each peak, for the EMDE scenario of Fig.~\ref{fig:histcomp} (defined by $m_Y = 4.5$ TeV, $\trh = 15.2$ MeV and $\eta = 1000$). The bulk of the annihilation signal is associated with peaks that are predicted to collapse after $0.3 \aeq$ (vertical dotted line).
}
\label{fig:jwac}
\end{figure}

Figure~\ref{fig:histcomp} shows that even after accounting for the suppression caused by gravitational heating, enough density fluctuation power can remain that a majority of the peaks are predicted to collapse before matter-radiation equality.
We wish to restrict our analysis to cases in which structure does not begin to form too deep in radiation domination, due to modeling uncertainties at such times.
However, for the scenario in Fig.~\ref{fig:histcomp}, the vast majority of the peaks are predicted to collapse after $0.1\aeq$, late enough that they should be locally matter dominated by the time of collapse \cite{2023MNRAS.520.4370D,2023PhRvD.107h3505D}. Moreover, Figure \ref{fig:jwac} -- which plots a $J$-weighted histogram of the collapse times for the same scenario as Fig.~\ref{fig:histcomp} -- shows that the bulk of the annihilation signal comes from peaks collapsing after $0.3\aeq$,\footnote{For peaks predicted to collapse long before matter domination, our calculation tends to predict large core sizes, which suppresses how much these peaks contribute to the annihilation signal in our analysis. This outcome is likely conservative (see Appendix~\ref{app:invcore}).} a late enough time that they would be locally matter dominated long before collapse, and so their collapse can be expected to proceed almost the same as during the matter era. 
For our main results in Section~\ref{results}, we will restrict our consideration to cosmologies in which structure formation begins no earlier than in this example case. In particular, for the scenario in Figs. \ref{fig:histcomp} and~\ref{fig:jwac}, Press-Schechter theory predicts that about 10\% of the dark matter is in bound structures by $a=0.3\aeq$. We will restrict our analysis to cosmologies for which Press-Schechter prediction is that less than 10\% of the dark matter is bound by $a=0.3\aeq$.

\subsection{Impact of Stellar Encounters and Tides}
\label{cusp_surv}

Prompt cusps persist as substructure inside current-day halos. Mergers between halos' central cusps can reduce their counts, and we account for this effect in our final analysis (explained in Sec.~\ref{results}). In addition, matter inside substructure cusps is stripped away by tidal forces and encounters with stars as they orbit inside their parent halos. In this section, we study the impact of these disruptions on the annihilation signal of EMDE cusps. 

The \textsc{cusp-encounters} code models the effect of a stellar encounter on a cusp using the quadrupolar tidal field of a passing star. A star of mass $M_{*}$ passing by a cusp with relative velocity $v$ and impact parameter $b$ imparts a tidal shock of strength $B = 2GM_{*} / (vb^2)$ on the particles of the cusp, meaning that a particle at radius $r$ within the cusp receives a velocity kick $\Delta v\sim B r$ (relative to the velocity of the cusp center), although the precise kick depends on the three-dimensional position.
Comparing this $\Delta v$ to the circular orbit velocity $v_{\rm circ}$ within the cusp, i.e. comparing $B$ to $v_{\rm circ}/r$, provides a guideline for the radii from which particles are expected to be ejected.
At the cusp and core radii, we have \begin{align}
B_{\rm cusp} & \equiv \frac{v_{\rm circ}(\rcusp)}{\rcusp} = \sqrt{\frac{8 \pi G A}{3 \rcusp^{3/2}}} , \\
B_{\rm core} &\equiv \frac{v_{\rm circ}(\rcore)}{\rcore} = \sqrt{\frac{8 \pi G A}{3 \rcore^{3/2}}} \propto A^{2/3} \fmax ^{1/3}.
\end{align} Shocks with $B > B_{\rm cusp}$ eject particles at $\rcusp$ (and similarly with $B_{\rm core}$). These two $B$ values are therefore indicative of the resilience of a cusp to tidal disruptions. For instance, shocks with $B > B_{\rm core}$ might completely disrupt a cusp. In this context, it is instructive to point out that $B^2/G \sim$ the density at a given cusp radius.

\begin{figure}[h!]
\centering
\includegraphics[width=0.95\textwidth]{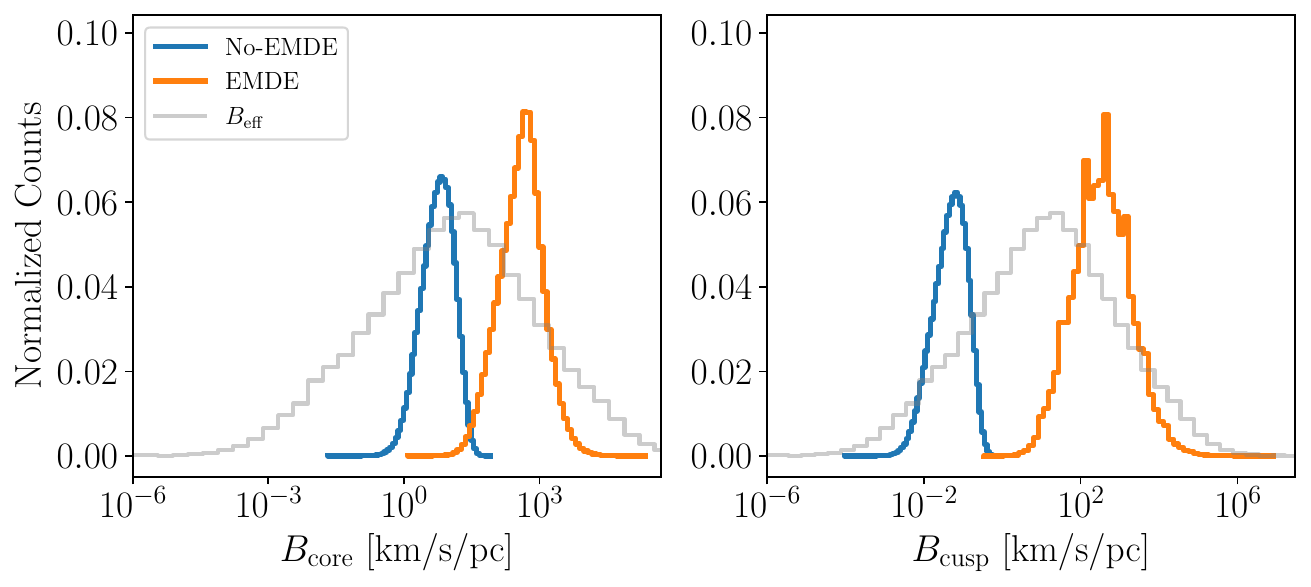}
\caption{Histograms of the shock strength $B$ required to disrupt the particles at the core radii (left) and cusp radii (right) for the no-EMDE (blue) and EMDE (orange) cusps. The grey histogram shows the distribution of $B_{\rm eff}$, the accumulated shock strength imparted by stellar encounters, for a collection of orbits in a Milky Way potential. We simulated $10^5$ orbits, out of which the 78\% that experienced stellar shocks are shown here. We consider the same EMDE cosmology as Fig.~\ref{fig:histcomp} with $m_Y = 4.5$ TeV, $\eta = 1000$ and $\trh = 15.2$ MeV.}
\label{fig:bcomp}
\end{figure}

Figure~\ref{fig:bcomp} shows histograms of $B_{\rm core}$ and $B_{\rm cusp}$ for the no-EMDE (blue) and EMDE (orange) cusps described in Figure~\ref{fig:histcomp}. As discussed previously, $A$ is much larger and $\rcore$ and $\rcusp$ much smaller for the EMDE cusps, resulting in higher $B$ values at the cusp and core radii and making these cusps much more resistant to mass stripping due to stellar encounters.

We used the \textsc{cusp-encounters} code to deform the annihilation signal of sampled cusps to assess the impact of these disruptions. The code models the stellar and gas disks, the stellar bulge, the baryon population, and the halo of the Milky Way using a variety of models \cite{cuspe}. In then integrates the orbits of an ensemble of test particles in the potential of this Milky Way model. For each orbit, the code creates a history of stellar encounter shocks as a series of $B$ values and models their cumulative effect as $B_{\rm eff}$, the 1.2-norm of this $B$-vector. We use the same set of $10^5$ orbits as in Ref.~\cite{cuspe}, obtained from their public code repository. The grey histograms in Figure \ref{fig:bcomp} show the distribution of $B_{\rm eff}$.
Most of the $B_{\rm core}$ values for the no-EMDE cusps are lower than the $B_{\rm eff}$ at which the distribution peaks, indicating that many of these cusps will be fully disrupted along their orbits. On the other hand, the $B_{\rm core}$ values for the EMDE cusps are mostly higher than the $B_{\rm eff}$ at which the distribution peaks, implying that many inner cores of the EMDE cusps will remain intact as they orbit in the Milky Way halo. 

The code also incorporates the impact of gradual tidal stripping due to the tidal field of the host by modifying $B_{\rm eff}$ to $B_{\mathrm{eff}, \lambda} = (B_{\rm eff}^2 + 42.2 \lambda^2)^{1/2}$, where $\lambda$ is the largest eigenvalue of the tidal tensor of the angle-averaged mass distribution of the Galaxy and its halo at the cusp orbit pericentre. The code modifies the J-factor of each cusp using a fitting function that depends on the $B_{\mathrm{eff}, \lambda}$ of its orbit. 

\begin{figure}[h!]
\centering
\includegraphics[width=0.95\textwidth]{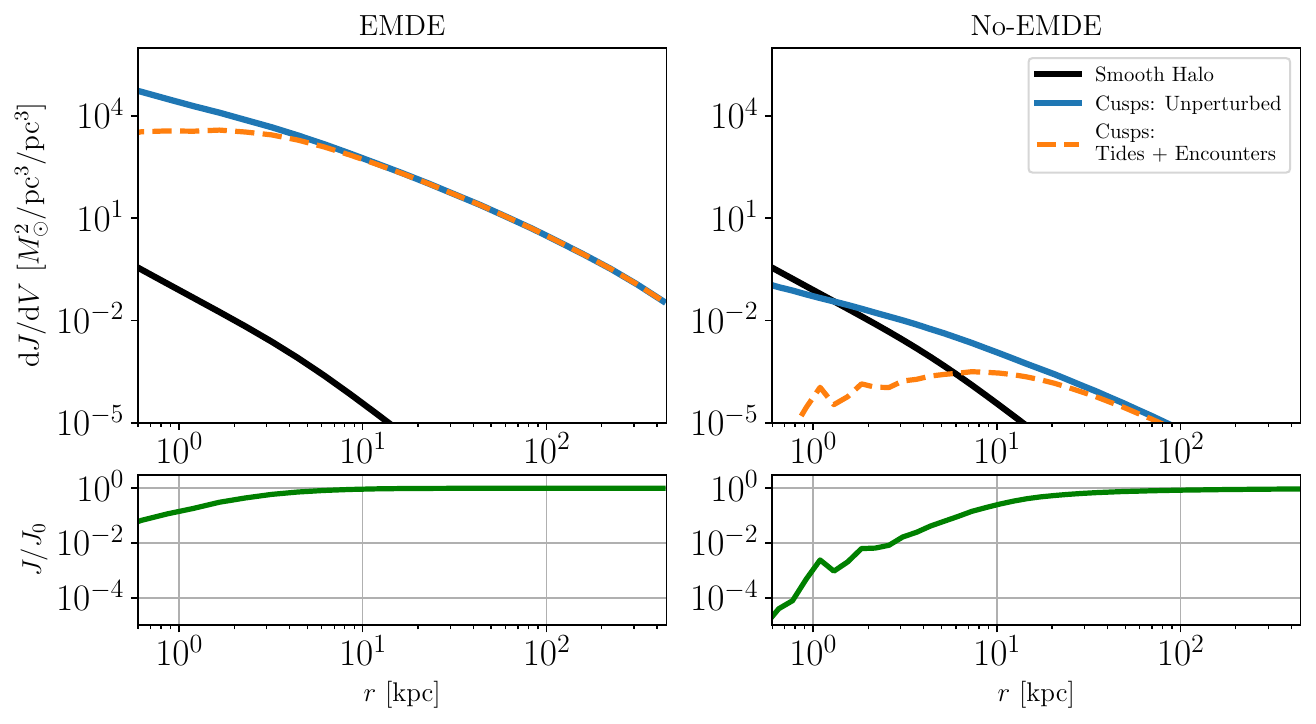}
\caption{\textbf{Top}: The distribution of the annihilation signal per unit volume in spherical shells at radius $r$ from the centre of a Milky Way halo, comparing the no-EMDE $\Lambda$CDM (right) and EMDE (left) cases. The EMDE scenario corresponds to $m_Y = 4.5$ TeV, $\eta = 1000$ and $\trh = 15.2$ MeV, as in previous figures.
The blue curves show the contribution from the cusps if the disruptive effects are neglected, whereas the dashed orange lines show the same contribution accounting for stellar encounters and tidal stripping. \textbf{Bottom}: the ratio of the blue and orange curves as a function of $r$, showing the suppression of the annihilation flux due to stars and tides.}
\label{fig:jradial}
\end{figure}

In this way, a radial profile of the suppression of the J-factor in the host halo is obtained, shown in Figure~\ref{fig:jradial}. The different curves in the top panels show the total $J$ in each volume bin as a function of the radius $r$ from the centre of the Milky Way halo. The blue curves show the $J$ from cusps, the dashed orange curves show the $J$ after suppression due to stellar encounters and tides, and the black curves show the $J$ contributed by the smooth halo without substructure. The cusp contribution to the $J$ in each bin is much higher in the EMDE case (left) than in the case without an EMDE (right), owing to the higher central densities of the cusps in the former. The lower panels show the suppression as $J/J_0$, the ratio of the orange to blue curve values for each $r$. The suppression is much weaker and becomes significant at much smaller radii for EMDE cusps as a result of their resilience to disruption. In contrast, the no-EMDE cusps for $r < 10$ kpc are heavily impacted, significantly reducing the annihilation signal at those radii. 

\begin{figure}[h!]
\centering
\includegraphics[width=0.95\textwidth]{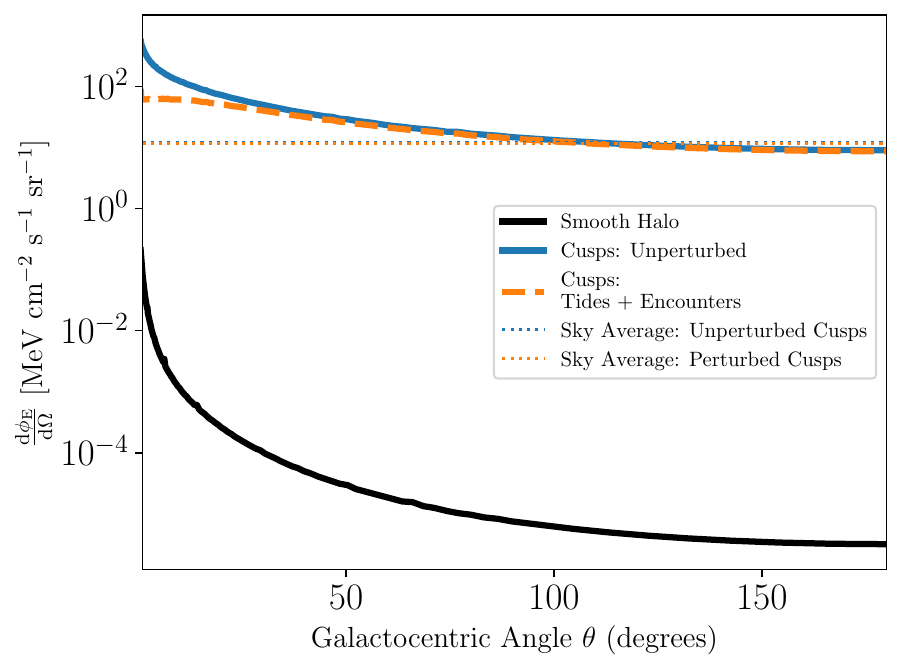}
\caption{The DM annihilation flux from a population of EMDE-induced cusps inside the Milky Way halo as a function of the Galactocentric angle. The blue solid curve shows the flux from unperturbed cusps, while the dashed orange curve shows the flux with the effects of tidal stripping and stellar encounters added. The dotted lines show the sky averages of the two angular profiles. As in previous figures, the EMDE scenario corresponds to $m_Y = 4.5$ TeV, $\eta = 1000$ and $\trh = 15.2$ MeV.
}
\label{fig:jang}
\end{figure}

To assess how this effect changes the annihilation signal from cusps within our Galactic halo, we used the code to generate a profile of the observed annihilation flux as a function of Galactocentric angle $\theta$ both with and without the disruptive effects. The code integrates the flux over the line of sight. This angular annihilation profile for a population of EMDE cusps from the same cosmology as in Figure~\ref{fig:histcomp} is shown in Figure~\ref{fig:jang}, with the solid blue curve showing the flux from cusps without the effects of tidal stripping and stellar encounters and the dashed orange curves showing the same after those effects are incorporated. We averaged the two over a large random sample of $\theta$, limiting the range to Galactic latitudes $|b| > 20$ degrees to match the cut of the LAT data (as in Ref.~\cite{cuspe}): these are shown by the dotted lines. The ratio of the two sky averages is greater than 0.96, indicating that the suppression of the annihilation signal is a sub-5\% effect for EMDE cusps. Although this suppression can vary between EMDE scenarios, we will neglect it hereafter since the EMDE cusps are much denser and more resistant to disruption in general than $\Lambda$CDM cusps. We will also neglect the much smaller aggregate impact of tidal stripping and stellar encounters on extragalactic prompt cusps.

\section{Using the IGRB to Constrain EMDEs}
\label{results}

The Large Area Telescope (LAT) operated by the Fermi Collaboration has detected a diffuse background of $\gamma$ rays of energies ranging from 100 MeV to 820 GeV \cite{fermilat}. It is now known that this isotropic $\gamma$-ray background (IGRB) is dominated by emissions from astrophysical sources including blazars \cite{blazar_igrb}, radio galaxies \cite{linden_radio}, active galactic nuclei \cite{inoue}, and star-forming galaxies \cite{sfgal_igrb,blanco_sfgal}. Apart from these sources, the products of the annihilation or decay of DM particles from the Milky Way halo and outside the Galaxy could also contribute to this background. 

The attribution of most of the IGRB signal to known sources leaves little room for a dark matter-sourced component, allowing constraints to be placed on dark matter properties (e.g. \cite{dimauro15}). For instance, Ref.~\cite{sten_cusps_dma} recently modeled the dark matter annihilation from prompt cusps in a $\Lambda$CDM scenario, using the IGRB to constrain the dark matter annihilation cross-section. As another example, Ref.~\cite{decaydm} modeled the contribution of decaying dark matter to the IGRB and set limits on the decaying dark matter lifetime. 

Since the dark matter annihilation rate
is dominated by prompt cusps, the annihilation signal from a volume is expected to track the number of cusps in that volume and hence the average dark matter density, as opposed to the density squared. As a result, the morphology of the signal in such scenarios is like that resulting from decaying dark matter. We can use this similarity to rescale the constraints on the decaying DM lifetime obtained in Ref.~\cite{decaydm} to constrain the DM annihilation cross-section in EMDE cosmologies. 

Considering a mass $M$ of dark matter which contains a large number of prompt cusps, the total volume integral of the density squared arising from the cusps, per unit mass, is \begin{equation}
    \frac{J}{M} \equiv \frac{\int \rho^2 \mathrm{d}V}{M} = \frac{\sum_{i} J_{i} }{M} = \frac{n_{\rm pk} \langle J \rangle}{\rho_0},
\end{equation} where $n_{\rm pk}$ is the comoving number density of initial density peaks, $\rho_0$ is the background DM density today, and angle brackets denote averaging over the sample of cusps. 

For each EMDE scenario, we use our modified \textsc{cusp-encounters} code to sample $N = 10^5$ peaks from the power spectrum and obtain the $J/M$ from the resultant cusps. If the fraction of mass in halos predicted by Press-Schechter theory exceeds 1\% at the end of the EMDE (about the smallest fraction considered in the simulations of Ref.~\cite{heating}), we apply the gravitational heating-caused free-streaming cut-off on the power spectrum. For all other cases, the EMDE-enhanced power spectrum without gravitational heating is used to sample the peaks. As we discussed in Section~\ref{code}, we also restrict our analysis to cosmologies for which Press-Schechter theory predicts that less than 10\% of the dark matter is bound into structures by $a=0.3\aeq$. By excluding a range of intermediate cases, this restriction makes a clear separation between the gravitationally heated and not-heated scenarios.

\begin{figure}[h!]
\centering
\includegraphics[width=\textwidth]{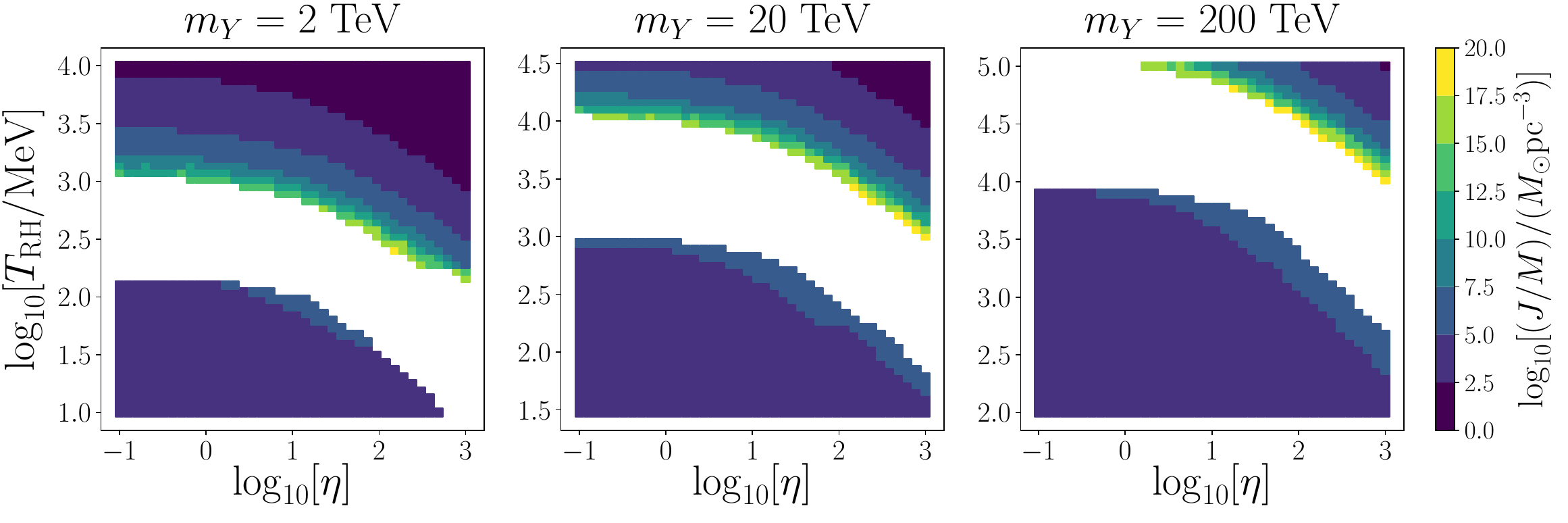}
\caption{Values of the DM annihilation J-factor per unit DM mass, $J/M$, shown for EMDE cosmologies with three different values of $m_Y$, as functions of reheat temperature $\trh$ and $\eta$, defined as $\rho_{\rm SM} / \rho_Y$ when the $Y$ particles are relativistic. The values at the bottom show the heating cases, while the top values show the no-heating cases. Note that in the heating regime, $J/M$ remains above $10^{3}~M_\odot\mathrm{pc}^{-3}$ for all cases shown.
} 
\label{fig:jmconts}
\end{figure}

Figure~\ref{fig:jmconts} shows the contours of $J/M$ for EMDEs caused by $Y$ particles of three different masses, $m_Y=2$, 20, and 200 TeV, as functions of the reheating temperature $\trh$ and $\eta$, which is the ratio $\rho_{\rm SM} / \rho_Y$ at a time when the $Y$ particles are relativistic. The duration of the EMDE generally decreases in the direction of increasing $\trh$ and $\eta$. For the bottom contours, which show cases with gravitational heating, a shorter EMDE translates to a larger power spectrum bump after heating, which means a higher $J/M$ value. The contours on the top show the no-heating cases. Without the EMDE enhancement to the power spectrum suppressed by the heating cut-off, a shorter EMDE means a smaller power spectrum bump, leading to lower $J/M$ values. Comparing the no-heating contours across plots with different $m_Y$ values, it is also apparent that a higher $m_Y$ generally results in a higher $J/M$ for the same $\eta$ and $\trh$. This is because a heavier $Y$ particle becomes cold earlier, leading to less suppression of the EMDE-enhanced power spectrum due to its relativistic pressure. In contrast, the $J/M$ values for the heating cases show comparatively less variation with $m_Y$ since heating erases the effect of the relativistic pressure of the $Y$ particles.
The heating-processed power spectrum always has a maximum value close to unity at matter-radiation equality \cite{heating}.

Given a value of $J/M$, the annihilation rate per DM mass is \begin{equation}\label{annihilation_per_mass}
      \frac{\Gamma}{M}  = \frac{\cs}{2m_X^2} \frac{J}{M} f_s.
\end{equation} In this equation, the factor $f_s$ accounts for the mergers of cusps, because of which not all collapsed peaks survive as cusps to the current day. Following Ref.~\cite{sten_cusps_dma}, we take this factor to be 0.5.
We show in Appendix~\ref{sim-survival} that this is a conservative estimate for EMDE cosmologies. 
As was discussed in Sec.~\ref{cusp_surv}, EMDE cusps are mostly resistant to stripping by stellar encounters and tidal forces and experience negligible reduction of their annihilation signal, so we neglect the impact of these phenomena. 

Under the assumption that DM annihilation and decay produce two primary particles, their rates of particle production can be equated to relate the DM annihilation signal to the decaying DM lifetime. 
A mass $M$ of annihilating DM with particle mass $m_X$ produces $2M(\Gamma/M)$ $Y$ particles, or $4M(\Gamma/M)$ SM particles, per unit time, while the same mass of decaying DM with particle mass $m_X$ and lifetime $\tau$ yields SM particles of the same energy at the rate $(2 / \tau) [M / m_X]$. By setting these rates to be the same, we find that the effective DM lifetime for annihilating DM is $\tau_{\rm eff} = 1/[2m_X (\Gamma/M)]$. Employing this conversion, lower limits on $\tau$ translate to upper limits on the DM annihilation cross-section $\cs$. 

\begin{figure}[h!]
\centering
\includegraphics[width=0.95\textwidth]{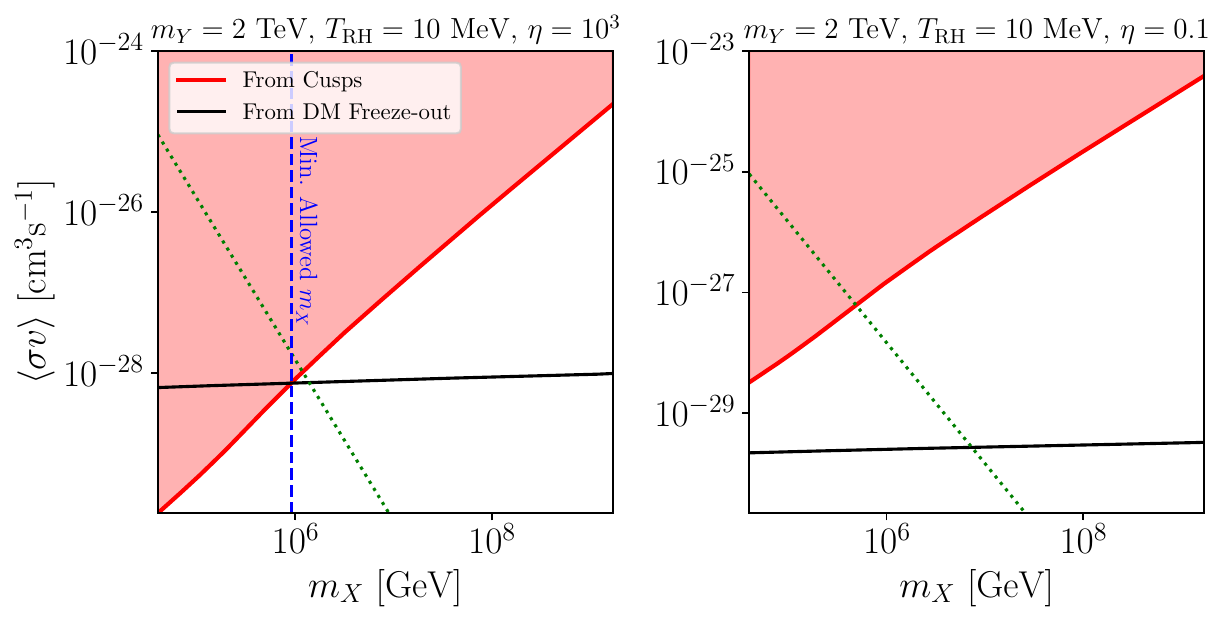}
\caption{Constraints on the dark matter annihilation cross-section for two EMDE cosmologies with the gravitational heating prescription applied. The red lines show the upper bounds on $\cs$ as a function of DM mass $m_X$. The black lines show the theoretical $\cs$ required for the correct DM relic abundance in that EMDE cosmology. The intersection of the red and black lines in the left panel defines the lower bound on the allowed $m_X$ for that EMDE case, shown by the blue dashed line. In the right panel, the upper bound on $\cs$ does not intersect the theoretical $\cs$ for any $m_X$, indicating that all $m_X>20m_Y$ are allowed for this cosmology. The green dotted lines show $\cs = 4 \pi / m_X^2$ {\textemdash} the unitarity bound is violated for $\cs$ above these lines.}
\label{fig:svp_heat}
\end{figure}

Figure \ref{fig:svp_heat} shows the upper limit on $\cs$ as a function of dark matter mass $m_X$ (red curve) translated from the lower limits on the decaying DM lifetime $\tau$ considering the $\tau^+ \tau^-$ decay channel from Ref.~\cite{decaydm}.
In the left panel, we consider the EMDE cosmology defined by $m_Y = 2$ TeV, $\rh{T} = 20$ MeV and $\eta = 10^3$. We impose the gravitational heating cut-off on the power spectrum since 29\% of the matter is bound in structures at the end of the EMDE. The cusp sampling yields $J/M = 50072$ $M_{\odot}/\textrm{pc}^3$. We consider a range of $m_X$ starting from $20m_Y$, because $m_X \gg m_Y$ is required for a hidden sector \cite{hid7}. The black line shows the $\cs$ value required for the correct DM abundance, calculated using our analytical treatment of DM freeze-out (outlined in Appendix \ref{app:dmfo}). The black line intersects the red curve, defining the lower limit on the allowed $m_X$, which is 912.17 TeV (shown by the blue dashed vertical line). 

The right panel of Figure~\ref{fig:svp_heat} shows a similar plot for a case with the same $m_Y$ and $\rh{T}$ but with $\eta = 0.1$. This case also has a heating cut-off on the power spectrum. Since $\eta$ is much lower compared to the case in the left panel, the power spectrum at the end of the EMDE has a much taller bump. This results in a higher velocity dispersion and thus a larger free-streaming scale. Consequently, this case yields a lower power spectrum bump after the EMDE and therefore cusps of lower density, generating $J/M = 2823$ $M_{\odot}/\textrm{pc}^3$. A lower $J/M$ implies a higher upper bound on $\cs$, and the black and red lines do not intersect. Therefore, the IGRB permits all $m_X > 20m_Y$ for this model. Finally, the dotted green lines in both panels of Figure~\ref{fig:svp_heat} show $\cs = 4\pi / m_X^2$; points in the plots above these lines violate the unitarity limit $\cs < 4\pi / m_X^2$ \cite{1990PhRvL..64..615G}. In both cases, the minimum $m_X$ consistent with the IGRB is also allowed by the unitarity condition.

In cases for which gravitational heating is considered, the core sizes are set by the linear velocity dispersion derived from the power spectrum at the end of the EMDE. They depend only on $m_Y$, $\rh{T}$ and $\eta$. On the other hand, when the heating prescription is not applied, the core sizes are set by the product of the scale factor and the DM velocity at freeze-out. This combination, and thus the $J/M$, depends on $m_X$. For scenarios without gravitational heating, we sample the peaks for ten $m_X$ values spaced logarithmically from $20m_Y$ to $2 \times 10^5 m_Y$ for each EMDE cosmology. We then interpolate between these ten values of $J/M$ to obtain the upper limit on $\cs$ as function of $m_X$. 

\begin{figure}[h!]
\centering
\includegraphics[width=0.95\textwidth]{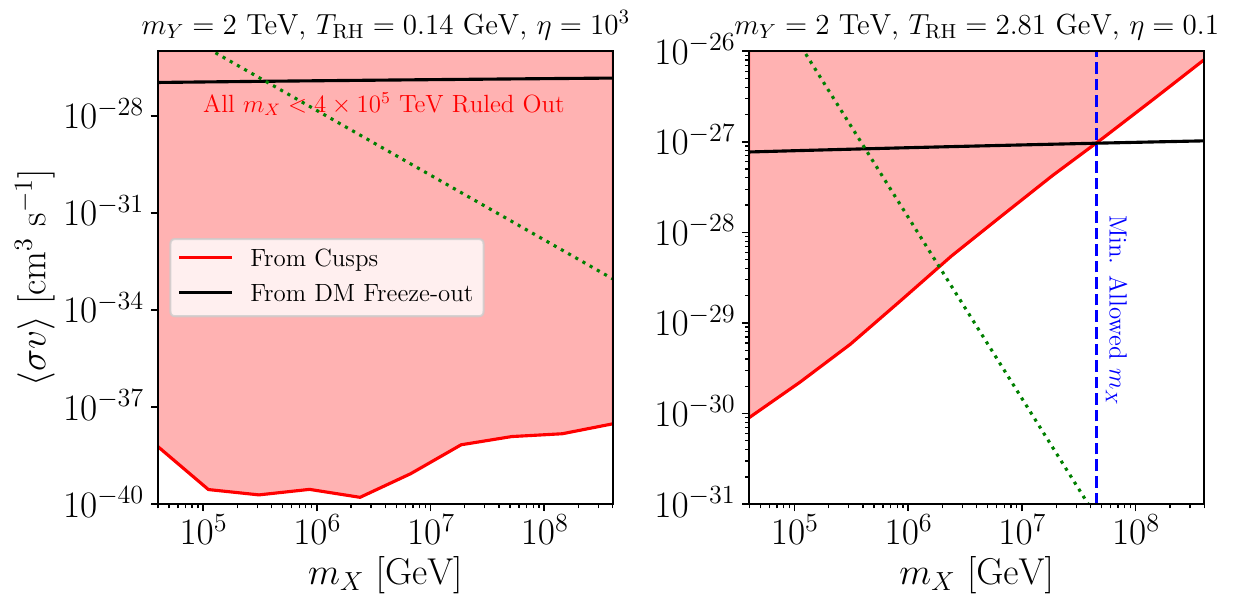}
\caption{Constraints on the DM annihilation cross-section for two cosmologies without the heating cut-off (c.f. Figure~\ref{fig:svp_heat}). In the left panel, the upper bound on $\cs$ (red curve) is lower than the $\cs$ required from DM freeze-out (black line), ruling out all $m_X$ values from $20m_Y$ to $2 \times 10^5 m_Y$. In the right panel, the red and black lines intersect, defining a lower bound on $m_X$ for that EMDE case. The green dotted lines show $\cs = 4 \pi / m_X^2$ {\textemdash} the unitarity bound is violated for $\cs$ above these lines.}
\label{fig:svp_noheat}
\end{figure}

Figure \ref{fig:svp_noheat} shows two such cases. In the left panel, we show the $\cs$-$m_X$ constraint plot for an EMDE with $m_Y = 2$ TeV, $\rh{T} = 0.14$ GeV and $\eta = 10^3$. As in Fig.~\ref{fig:svp_heat}, the black line shows the $\cs$ required for the correct DM abundance. For $m_X$ ranging from $40$ to $4 \times 10^5$ TeV, the black line lies above the red curve which shows the upper bound on $\cs$ for this case {\textemdash} all $m_X$ in this range are ruled out. The right panel shows an EMDE case in which a lower bound on $m_X$ is obtained. However, the $m_X$ obtained by the intersection of the red and black curves violates the unitarity limit represented by the green dotted line {\textemdash} this EMDE scenario is ruled out, at least for point-like dark matter particles.

\begin{figure}[h!]
\centering
\includegraphics[width=0.95\textwidth]{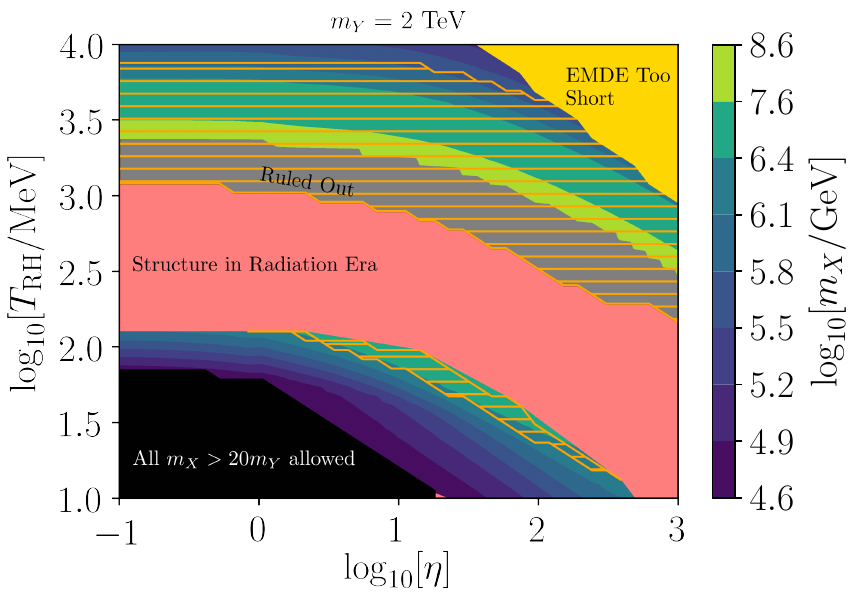}
\caption{Lower bound on the allowed DM mass ($m_X$) due to the IGRB, considered as a function of $\trh$ and $\eta$ (defined as $\eta \equiv \rho_{\rm SM} / \rho_Y$ when the $Y$ particles are still relativistic). This plot covers EMDE cosmologies with $m_Y = 2$ TeV.
Scenarios with gravitational heating due to a long EMDE lie below the red band, while scenarios in which the first nonlinear structures form during or close to the late matter epoch lie above the red band. Scenarios within the intermediate red band are not accurately represented by our models and require further study, although they are expected to be tightly constrained \cite{blanco19}.
In the black region, all $m_X > 20m_Y$ are consistent with the IGRB.
In the grey region, the lower limit on $m_X$ is extremely high and exceeds the maximum of the color scale.
Within the hatched regions, the IGRB rules out all combinations of $m_X$ and $\cs$ that generate the observed dark matter abundance through freeze-out and respect the unitarity limit.
}
\label{fig:m2mx} 
\end{figure}

The intersection (or non-intersection) of the upper bound on $\cs$ with the required $\cs$ thus translates the $\cs$-$m_X$ plot for each EMDE case to one value: the lower bound on the DM mass $m_X$ for that EMDE cosmology. This lower bound can be mapped on to the parameter space of EMDEs. In Figure \ref{fig:m2mx}, we show the parameter space for $m_Y = 2$ TeV by plotting contours of the constraints on $m_X$ as a function of $\rh{T}$ and $\eta$.
Here, and for the remainder of this work, we follow previous works (e.g.~\cite{blanco19,sten_gr}) in assuming that the annihilation product is $b\bar{b}$.\footnote{That is, we convert from limits on dark matter decaying into $b\bar b$. The spectrum of gamma rays produced by dark matter annihilation that proceeds through an intermediate $Y$ state is very similar to the spectrum from dark matter decay into the same final state, with only a small sensitivity to the mass of the intermediate state \cite{hid3}.} Different decay channels generally shift the limits by less than a factor of two \cite{decaydm}.
Additionally, the subsequent contour plots all consider a bosonic mediator particle causing the EMDE. In the case of a fermionic mediator, the general trend of these plots remains the same. However, for a fermionic $Y$, the maximum of the bump in the EMDE-processed power spectrum is at a larger scale
by about 11\%
compared to the boson case for the same EMDE parameters.
This reduces the bump height, leading to slightly later structure formation. Consequently, for the no-heating cases, $J/M$ is decreased by a factor of about 2, and the bound on $m_X$ is weakened by a comparable factor. For the heating cases, reduced structure formation during the EMDE logarithmically raises $J/M$ and thus modestly strengthens the bound on $m_X$.

The contours on the bottom in Figure~\ref{fig:m2mx} show the heating cases while those on the top show the no-heating cases.
The red region marks the intermediate parameter range that violates our restriction that $<10\%$ of the dark matter be in bound structures by $a=0.3\aeq$ (according to Press-Schechter theory).
Those scenarios are expected to yield large annihilation signals \cite{blanco19}, but more work is needed to understand them precisely.
It is notable that the contours (both heating and no-heating) become independent of $\eta$ for $\eta \ll 1$. For this range of $\eta$, the SM radiation density is subdominant to $\rho_Y$ before the EMDE and thus has no effect on the cosmological background or the matter power spectrum. Consequently, the exact value of $\eta$ has no effect on our results for $\eta \ll 1$.

The black region below the heating contours shows the parameter space for which all $m_X > 20m_Y$ are consistent with the IGRB, an example of which was shown in the right panel of Figure~\ref{fig:svp_heat}. The grey region neighboring the no-heating contours covers the cases in which all $m_X$ from $20m_Y$ to $2 \times 10^5 m_Y$ are ruled out by the IGRB for the no-heating cases, an example of which was shown in the left panel of Figure~\ref{fig:svp_noheat}.

Increasing $\rh{T}$ or $\eta$ generally shortens the EMDE. Recall that $\eta$ is defined as $\rho_{\rm SM} / \rho_Y$ when the $Y$ particles are still relativistic (but the DM is cold). A larger $\eta$ therefore causes $\rho_Y$ to start dominating over $\rho_{\rm SM}$ later. Similarly, a larger $\rh{T}$ means earlier reheating.
For points in the gold region at very high $\eta$ and $\trh$, the EMDE is too short, i.e., $k_{\rm dom} / \krh < 3$, where $k_{\rm dom}$ and $\krh$ are the wavenumbers of modes entering the horizon at the start and end of the EMDE, respectively. Our EMDE transfer functions from Ref.~\cite{hg23} are not accurate in this regime.

For the cases with gravitational heating, a shorter EMDE means less structure formation before reheating and a lower velocity dispersion, which implies a shorter free-streaming length due to the heating. As a result, a shorter EMDE leaves a larger bump in the power spectrum after gravitational heating, generating a higher value of $J/M$. As was shown in Figure~\ref{fig:svp_heat}, a higher $J/M$ translates into a higher value of the lower bound on $m_X$. This effect is apparent in the contours for the heating cases in Figure~\ref{fig:m2mx}: the contour values increase in the direction of increasing $\eta$ and $\trh$. However, this trend is also driven by variation of the $\cs$ needed to reach the known dark matter abundance. As we will explore later, a longer EMDE leads to a smaller required $\cs$.

\begin{figure}[h!]
\centering
\includegraphics[width=0.95\textwidth]{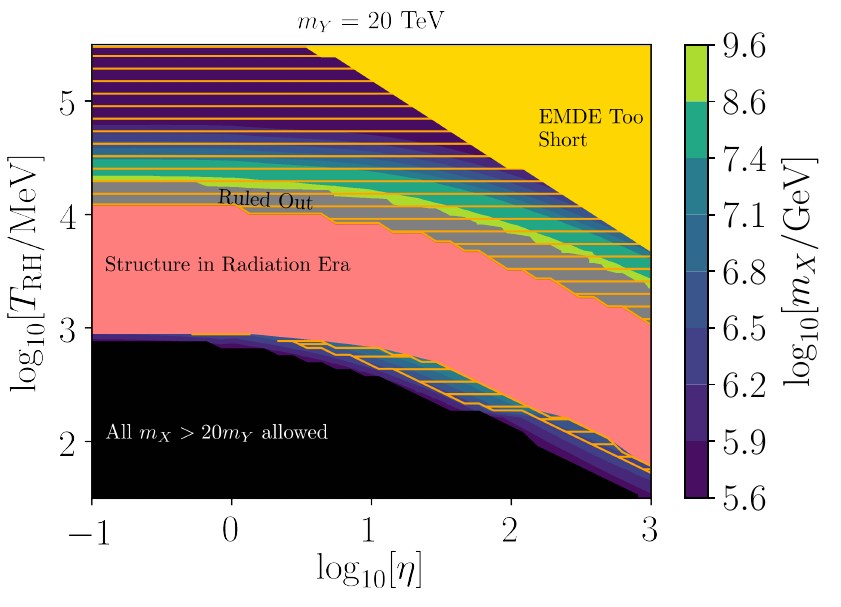}
\caption{As in Figure~\ref{fig:m2mx}, but for EMDEs with $m_Y = 20$ TeV.}
\label{fig:m20mx}
\end{figure}

For the no-heating cases, a shorter EMDE implies a smaller enhancement to the power spectrum, which remains unprocessed by gravitational heating. In these cases, increasing $\rh{T}$ or $\eta$ shortens the power spectrum bump, lowering $J/M$ and weakening the constraint on $m_X$ as a result. The upper (no-heating) contours of Figure~\ref{fig:m2mx} show that the minimum allowed $m_X$ increases in the direction of decreasing $\rh{T}$ and $\eta$, a trend that goes opposite to that for the heating cases. Figure \ref{fig:m20mx} shows similar contours for EMDEs with $m_Y = 20$ TeV.

The orange hatched area in Figures \ref{fig:m2mx} and \ref{fig:m20mx} indicates where the lower limit on $m_X$ from the IGRB is high enough that it would make the dark matter violate the unitarity limit for point-like particles, $\cs<4\pi/m_X^2$.
Here $\cs$ is the annihilation cross section required for DM to freeze out at the known abundance (calculated according to Appendix~\ref{app:dmfo}).
The no-heating cases feature short EMDEs, which require larger $\cs$ to obtain the observed DM relic abundance. In conjunction with the IGRB limits, this makes $\cs>4\pi / m_X^2$ for most of the no-heating parameters, ruling them out. The required $\cs$ are much lower and the $m_X$ constraints much weaker for the heating cases (lower contours); much of this portion of the parameter space is therefore allowed. 

\begin{figure}[h!]
\centering
\includegraphics[width=\textwidth]{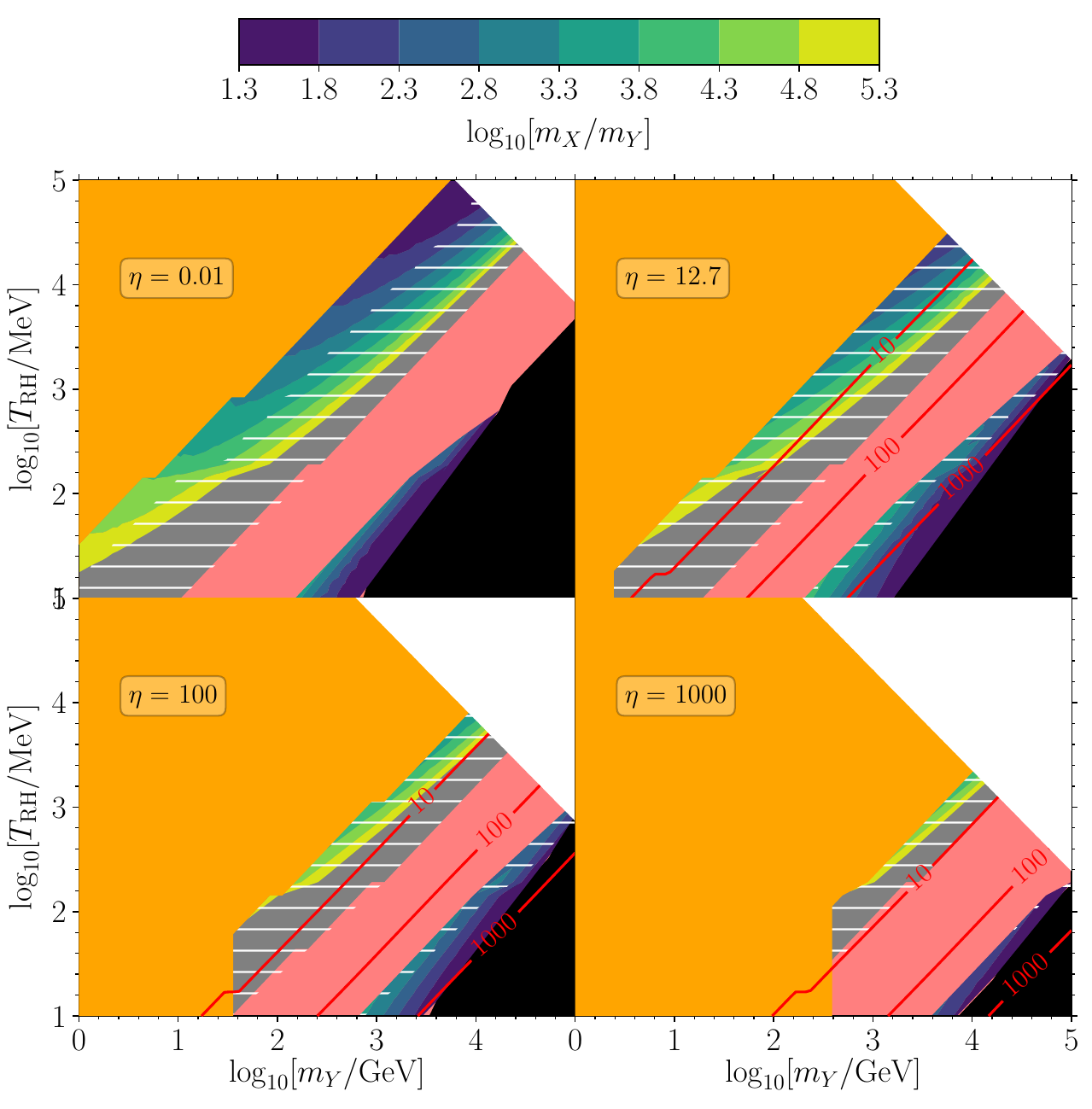}
\caption{Lower limits on $m_X$ (colors) due to the IGRB, shown as a function of $\trh$ and $m_Y$ for four different values of $\eta$.
The gold and red regions are not constrained by our analysis, with the gold region corresponding to a too-short EMDE and the red region corresponding to structure forming too deep in radiation domination (though we expect based on previous work that the red regions are ruled out \cite{blanco19}).
Below the red band in each panel lie scenarios with gravitational heating due to a long EMDE, while above it are scenarios without heating.
In the black areas, all $m_X > 20m_Y$ are allowed by the IGRB.
The grey band marks cases in which no $m_X < 2 \times 10^5 m_Y$ is allowed.
The white hatches show the parameters
for which all $m_X$ allowed by the IGRB violate the unitarity limit, while
the solid white area violates the unitarity limit for all $m_X > 20m_Y$.
Finally, the red lines show constant $T_{\rm dom}/\trh$ with the inline numbers showing the value.}
\label{fig:mxlimits}
\end{figure}

To cover a broader range of the mediator mass $m_Y$, we also show the lower limits on $m_X$ for four fixed $\eta$ values in Figure \ref{fig:mxlimits}. The upper left regions marked gold represent EMDEs that are too short ($k_{\rm dom} / \krh < 3$), while the red shaded regions show the parameter space with significant structure formation in the radiation-dominated era, as in Figures \ref{fig:m2mx} and \ref{fig:m20mx}.
The lower right contours show the heating cases, with all $m_X > 20m_Y$ allowed in the black regions. The diagonal band of contours in the middle of each panel shows the no-heating cases, with the grey regions covering the parameter combinations for which all $m_X < 2 \times 10^5 m_Y$ are ruled out.

Figure \ref{fig:mxlimits} also shows the parameter regions for which the unitarity condition is violated; 
the solid white regions violate unitarity for all $m_X>20m_Y$, while the white hatched regions violate unitarity for all $m_X$ consistent with the IGRB.
In each of these panels, the duration of the EMDE decreases with decreasing $m_Y$ and increasing $\trh$ (going towards the upper left). Similar to Figures \ref{fig:m2mx} and \ref{fig:m20mx}, the $m_X$ constraints strengthen as one moves to the upper left along the heating contours and weaken in the same direction along the no-heating contours. 

The three panels for cases with $\eta>1$ in Figure \ref{fig:mxlimits} also show diagonal red lines marking constant values of $T_{\rm dom} / \trh$, the ratio of the SM temperatures at the beginning and end of the EMDE. These lines enable a comparison with the results of Ref.~\cite{blanco19}. The EMDE cosmology chosen in Ref.~\cite{blanco19} corresponds to $\eta = 12.7$, shown in the top right panel. Our analysis, employing the exact power spectra for both heating and no-heating cases and the cusp formalism, rules out most EMDEs shorter than those with $T_{\rm dom} / \trh \lesssim 50$. Moreover, for $m_X = 20m_Y$, the white shaded region shows that a significant subset of the EMDE parameter space is ruled out by unitarity for $T_{\rm dom} / \trh \lesssim 700$. In comparison, Ref.~\cite{blanco19} only rules out cases for $T_{\rm dom} / \trh \lesssim 200$. 

\begin{figure}[h!]
\centering
\includegraphics[width=\textwidth]{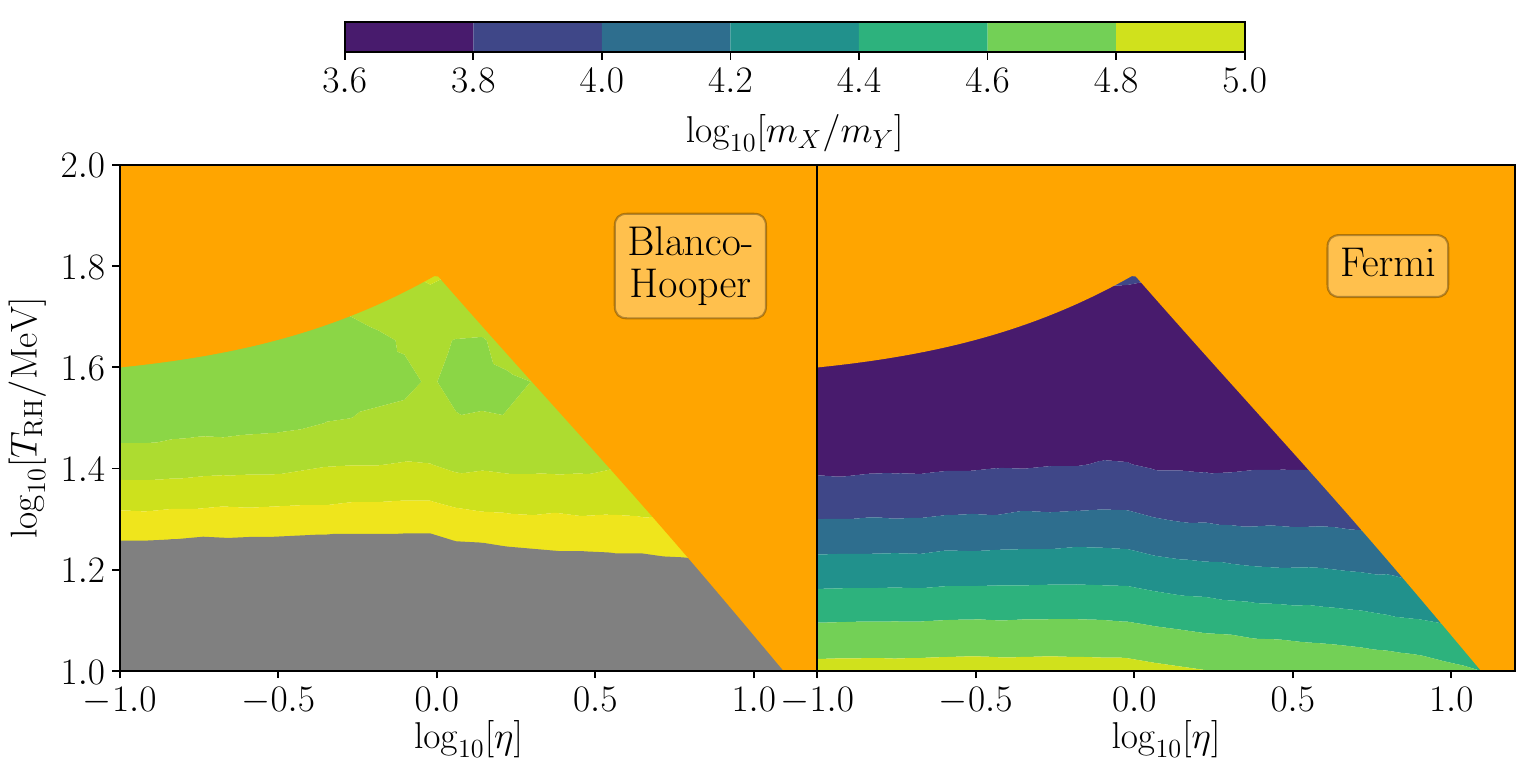}
\caption{Influence of the astrophysical foreground model on the lower limits on $m_X$ due to the IGRB. We consider the case of $m_Y=1$~GeV.
The left panel uses foregrounds from Ref.~\cite{decaydm} while the right panel uses foregrounds from Ref.~\cite{liu_fermi}. The orange regions mark EMDEs that are too short for our transfer functions to correctly model their power spectra ($k_{\rm dom} / \krh < 3$).}
\label{fig:foreground_compare}
\end{figure}

Our constraints have relied on the dark matter decay limits of Ref.~\cite{decaydm}, which used a model of astrophysical foregrounds that is based on multiwavelength observations and on the spatial distribution of the gamma-ray signal.
A simpler alternative, taken by Ref.~\cite{liu_fermi}, is to use an astrophysical foreground model that is fit by Ref.~\cite{igrb} to the measured IGRB alone.
To test the sensitivity of our results to the foreground modeling, Fig.~\ref{fig:foreground_compare} compares constraints obtained using the two approaches.
Because Ref.~\cite{liu_fermi} does not model electromagnetic cascades (e.g.~\cite{Blanco:2018bbf}), the limits therein extend only up to $m_X=10^5$~GeV. Therefore, since a hidden sector requires $m_X\gg m_Y$, we restrict our consideration to the $m_Y=1$~GeV case.
Evidently, the choice of foreground model changes the limits on $m_X$ by approximately an order of magnitude.

\section{Summary}
\label{summary}

Theories in which dark matter lives in a hidden sector can be difficult to constrain via direct detection or collider experiments, since these hidden sectors couple only very feebly to the Standard Model. Fortunately, many such theories feature an early matter-dominated era (EMDE), during which subhorizon matter perturbations are boosted on account of their linear growth with the scale factor. This enhancement leads to a bump in the matter power spectrum at small scales, due to which dense small-scale structures form much earlier than in scenarios without an EMDE. These dense objects can be detected via their gravitational and annihilation signatures, enabling the exploration of hidden sectors. 

We test hidden sector models with an EMDE by considering the contribution to the observed isotropic gamma-ray background (IGRB) from dark matter annihilation in prompt cusps.
These $\rho\propto r^{-3/2}$ cusps, which form from the collapse of smooth density peaks, have been recently shown to survive hierarchical halo growth and dominate the present-day annihilation signal.
We computed this signal by exploiting the tight link between prompt cusps and the linear matter power spectrum.
To accurately model the power spectrum resulting from an EMDE, we used recent prescriptions for the relativistic pressure of the EMDE-causing particles and for gravitational heating associated with the dissolution of structure formed during the EMDE.
We implemented our calculations within the framework of the \textsc{cusp-encounters} code \cite{cuspe}, and by using that code to simulate orbits in the Galactic halo, we found that stellar encounters and tidal stripping in the Milky Way only negligibly reduce (by $\lesssim 5\%$) the annihilation signal in EMDE cosmologies.
On account of the bump in the power spectrum, cusps in an EMDE cosmology form much earlier than in standard cosmologies, which makes them much denser and results in annihilation signals that are many orders of magnitude larger.

\begin{figure}[h!]
\centering
\includegraphics[width=0.7\textwidth]{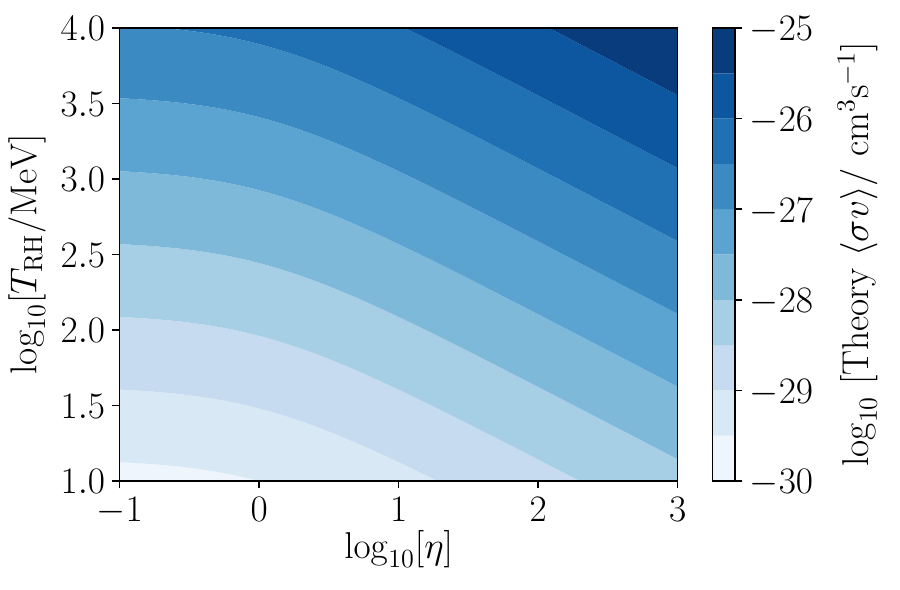}
\caption{Contours of the DM annihilation cross-section $\cs$ required for the correct DM abundance today, evaluated using the analytical method in Appendix~\ref{app:dmfo}, for EMDEs with $m_Y = 2$ TeV and $m_X = 100m_Y$. The duration of the EMDE generally increases in the direction of decreasing $\trh$ and $\eta$ (going toward the bottom left). Longer EMDEs require a smaller $\cs$ to obtain the observed DM relic abundance.}
\label{fig:svcont_m2}
\end{figure}

For each cosmological scenario, constraints from the IGRB can be framed in terms of a lower limit on the DM particle mass $m_X$, based on the requirement that the DM annihilation cross section $\cs$ be appropriate for it to freeze out at the known abundance.
These limits are shown in 
Figures~\ref{fig:m2mx}-\ref{fig:mxlimits}.
In the regime where gravitational heating is significant, shorter EMDEs lead to stronger constraints on the DM mass, while in cases where heating is not relevant, shorter EMDEs weaken the DM mass constraints.
Combined with the unitarity limit on the dark matter mass and cross section, our analysis fully rules out a large portion of the parameter range for hidden-sector-driven EMDEs, especially for high reheat temperatures $\trh$ and high masses $m_Y$ of the particle driving the EMDE.

The constraints on $m_X$ in the heating regime are much weaker than those in the no-heating cases. This difference is mostly because the longer EMDEs associated with the heating regime mean that $\cs$ must be lower to set the correct DM relic abundance. An example of this for $m_Y = 2$ TeV and $m_X = 100m_Y$ is shown in Figure~\ref{fig:svcont_m2}, which displays the required $\cs$ in the same parameter range as Fig.~\ref{fig:m2mx}. The smaller $\cs$ in the heating cases makes indirect detection more difficult, leading to weaker bounds on $m_X$.
It is noteworthy that despite poor prospects for the detection of annihilation radiation in the deep heating regime, the cusps forming in these scenarios are still very internally dense (e.g. Fig.~\ref{fig:jmconts}), so they remain promising targets for gravitational detection, especially since the low reheat temperatures of these scenarios are associated with the formation of more massive microhalos.
For instance, Ref.~\cite{pta_sten} showed that pulsar timing arrays could detect microhalos forming from EMDEs with $\trh \lesssim 150$ MeV.

We restrict our analysis to cases in which structure forms not too long before late-time matter domination. This restriction divides our parameter range into two regimes: one in which the EMDE is short and only moderately boosts density fluctuations, and the other in which abundant structure formation during a long EMDE gravitationally heats the dark matter.
Although the intermediate regime is expected to be tightly constrained \cite{blanco19}, further work is needed to precisely predict the annihilation signal arising from these scenarios. Our gravitational heating prescription does not describe them accurately enough, and our peak sampling methods are not fully accurate for collapse deep within radiation domination, since local matter domination is needed in addition to peak collapse to form a bound object during this epoch. 

Also, although we have assumed the DM annihilation signal to be isotropic for our analysis, the limits we have obtained could be tightened by using the anisotropy of the predicted signal from cusps in the Milky Way halo.
For example, Fig.~\ref{fig:jang} indicates that the predicted annihilation signal exceeds its sky-averaged value by nearly an order of magnitude for angles close to the Galactic Center. If astrophysical foregrounds can be kept under control in this regime, a comparable level of improvement on the limits on $\cs$, and hence $m_X$, might be possible.

\appendix

\section{Hidden Sector Freeze-out}
\label{app:dmfo}

In this appendix, we calculate the scale factor and temperature at which dark matter freezes out if it resides in a hidden sector with an early matter-dominated era (EMDE) caused by a light long-lived species ($Y$). This treatment is for cases with a bosonic $Y$ particle, but can easily be extended to cases with fermionic $Y$.

\subsection{Numerical Solution}

Consider a hidden sector at a temperature $\ths$ with a $Y$ particle with $g_Y$ degrees of freedom and dark matter $X$. Standard Model radiation resides in the visible sector. Dark matter annihilates to $Y$ particles in a reaction with a thermally averaged cross-section $\cs$, and the $Y$ particles decay into the visible sector with a decay rate $\Gamma$. The coupled system of equations for the densities of the $Y$ particles and the SM radiation along with the dark matter number density are: \begin{align} \label{dmfo_eq}
    \dot{\rho}_Y + 3H(1 + w)\rho_Y &= -\Gamma m_Y n_Y + \cs \langle E_X \rangle (n_X^2 - \nxe^2), \\
    \dot{\rho}_R + 4H\rho_{\rm SM} &= +\Gamma m_Y n_Y, \\
    \dot{n}_X + 3Hn_X &= \cs (n_X^2 - \nxe ^2),
\end{align} 
where overdots denote $d/dt$ and $H \equiv \dot{a}/a$. 
In the above equations, $\langle E_X \rangle \equiv \rho_X / n_X$ and $w \equiv P_Y / \rho_Y$ is the equation of state of the $Y$ particles, which controls their transition from relativistic to nonrelativistic behavior. We set $\langle E_X \rangle = (m_X^2 + 7.29\ths^2)^{1/2}$, following an approximation similar to that used in Ref.~\cite{ae15} but modified so that $\langle E_X \rangle$ matches the $\rho_X / n_X$ at high temperatures for a bosonic $X$ particle. In addition, we use a fit function from Ref.~\cite{hg23} to write\begin{equation}
    w(a) = \frac{1}{3} \left[ 1 + \left( \frac{am_Y}{3.05a_i T_{\rm hs,i}} \right)^{0.57} \right] ^{-\frac{1}{0.57}},
\end{equation} where $T_{\rm hs,i}$ is the hidden sector temperature at a time $a_i$ when the $Y$ particles are relativistic and the $X$ particles are nonrelativistic. Furthermore, the equilibrium number density of dark matter is \begin{equation}
    \nxe = \frac{g_X}{2\pi^2} \int_0^{\infty} \frac{p^2 dp}{\exp [\sqrt{p^2 + m_X^2}/T_{\rm hs}] - 1}.
\end{equation} The hidden sector temperature is set by the $Y$ particles. Following the calculation of the time evolution of $T_{\rm hs}$ in Ref.~\cite{hg23}, we use the following fitting function to model $\ths$: \begin{equation}
    T_{\rm hs}(a) = \frac{T_{\rm hs,i}}{a} \left[ 1 + \left( \frac{am_Y}{3.275a_i T_{\rm hs,i}} \right)^{0.62} \right] ^{-\frac{1}{0.62}}.
\end{equation}

We start our calculations at scale factor $a_i$ such that $T_{\rm hs,i} = 0.1m_X$, solving Eqs.~(\ref{dmfo_eq}) numerically to obtain $\rho_Y$, $\rho_{\rm SM}$ and $n_X$. We note here that our machinery for $\ths$ and $w_Y(a)$ and our definition of $\eta$ require us to begin calculations at a time when the $X$ particles are already nonrelativistic and the transfer of entropy to the $Y$ particles due to $X$ annihilations has become negligible. If we begin at an earlier time when the $XX \rightarrow YY$ process is significant, the comoving $Y$ particle density increases by a factor of $(1 + g_X / g_Y)^{4/3}$ via the entropy increase as $X$ particles annihilate into $Y$ particles. 

\begin{figure}[h!]
\centering
\includegraphics[scale=1]{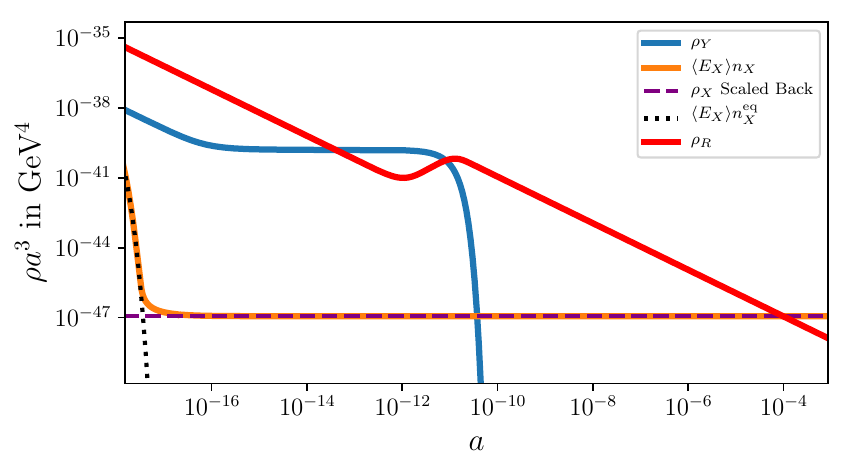}
\caption{\textbf{Hidden Sector DM Freeze-out in an EMDE cosmology}: Numerical solutions for the densities of the $Y$ particles (blue), the SM radiation (red), and the dark matter $X$ (orange) as a function of scale factor $a$. The black dotted curve shows the $\rho$ for the DM in equilibrium in the hidden sector. The DM freezes out when the orange curve dissociates from the black dotted curve. The purple dashed curve extrapolates the current-day DM density back in time by multiplying it with $a^{-3}$, and the orange curve matches it, indicating the DM has frozen out to yield the correct relic abundance.}
\label{fig:dmfo_full}
\end{figure}

\begin{figure}[h!]
\centering
\includegraphics[scale=1]{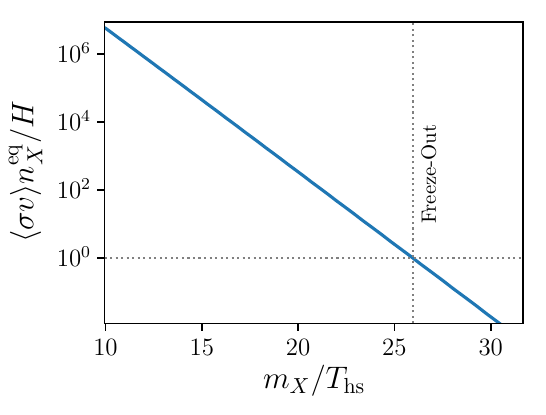}
\caption{The DM annhilation rate $\cs \nxe$ scaled by the Hubble rate $H$, shown as a function of $m_X / \ths$. Dark matter freezes out at the $\ths$ at which $\cs \nxe = H$. }
\label{fig:dmfo_x}
\end{figure}

To obtain the correct relic abundance of dark matter, we solved the equations with different values of $\cs$ for a given EMDE scenario. Figure \ref{fig:dmfo_full} shows $\rho a^3$ versus scale factor $a$ for the energy densities of the various fluids for an EMDE with $m_Y = 1$ TeV, $\eta = 500$ and $\trh = 20$ MeV. The dashed purple line shows the comoving dark matter density for the relic abundance of dark matter today. The black dotted line shows the evolution of $\langle E_X \rangle \nxe$, the equilibrium dark matter density. The orange curve shows $\langle E_X \rangle n_X$, the actual dark matter density which departs from the equilibrium density when the dark matter freezes out and matches the relic abundance. For this case, $\cs = 1.63 \times 10^{-11}$ GeV$^{-2}$ is needed to obtain the correct relic abundance. Figure \ref{fig:dmfo_x} shows the evolution of $\cs \nxe / H$ as a function of $x \equiv m_X / \ths$. Freeze-out occurs when $\cs \nxe = H$, which happens when $x = x_f = 25.92$.  

\subsection{Analytical Method}

Here we present an analytical calculation of the scale factor and dark matter temperature at freeze-out. We assume that the dark matter freezes out when it is nonrelativistic and the $Y$ particles are still relativistic. If the hidden sector temperature is $\ths$, we have $\rho_Y = g_Y (\pi^2 / 30) \ths^4$ and the SM radiation density $\rho_{\rm SM} = \eta \rho_Y$ (from the definition of $\eta$). With these, the Hubble rate is given by \begin{equation}
    H = \sqrt{ \frac{8 \pi G}{3} \rho_Y (1 + \eta)} = \sqrt{ \frac{8 \pi^3 G g_Y}{90} (1 + \eta)} \left[ \frac{m_X}{x} \right]^2, 
\end{equation} where $x \equiv m_X / \ths$ and $g_Y$ are the degrees of freedom of the $Y$ particle. In addition, the equilibrium number density of dark matter when it is nonrelativistic is \begin{equation}
    \nxe = g_X \left[ \frac{m_X^2 }{2 \pi x}\right]^{\frac{3}{2}} e^{-x}.
\end{equation}

The freeze-out condition is $H(x_f) = \cs \nxe (x_f)$ if freeze-out happens when $x = x_f$ at the scale factor $a_f$. The DM density at freeze-out is then \begin{equation} \label{focond}
     m_X \nxe (x_f) = \Omega_m \rho_{\rm crit} a_f^{-3},
\end{equation} where $\Omega_m$ is the matter density today in terms of the critical density $\rho_{\rm crit}$. Equating $\nxe$ from the two expressions above yields \begin{equation} \label{sigv_a}
    \cs = \frac{m_X H(x_f) a_f^3}{\Omega_m \rho_{\rm crit}} ,
\end{equation} where \begin{equation}
    H(x_f) = \sqrt{ \frac{8 \pi^3 G g_Y}{90} (1 + \eta)} \left[ \frac{m_X}{x_f} \right]^2.
\end{equation}

The scale factor $a_f$ can be rewritten in terms of the EMDE model parameters and $x_f$ using the background evolution model for $\rho_Y$ from Ref.~\cite{hg23}. When the $Y$ particles are relativistic, $\ths \propto a^{-1}$ such that $a \ths = a_p m_Y / 2.7$, where $a_p$ is the pivot scale factor at which $\rho_Y$ changes its behavior from going as $a^{-4}$ to $a^{-3}$. This relation implies $a_f T_f = a_p m_Y / 2.7$. Dividing both sides by $m_X$ and rearranging, we have \begin{equation}
   a_f = x_f \times \frac{m_Y}{m_X} \times  \frac{a_p}{2.7}. 
\end{equation} Using the expression for $a_p$ from G23, finally, \begin{equation} \label{af}
    a_f = 2.14 T_0 \left[ \frac{x_f}{m_X} \right] \left[ \frac{g_Y}{g_{*}(\trh) g_{*S}(\trh)}\right]^{\frac{1}{3}} \left[ \frac{\trh}{m_Y}\right]^{\frac{1}{3}},
\end{equation} where $T_0$ is the CMB temperature today, $g_{*}(T)$ is the number of relativistic degrees of freedom contributing to $\rho_{\rm SM}$ at temperature $T$ and $g_{*S}(T)$ is the number of degrees of freedom contributing to the entropy of the Standard Model radiation at temperature $T$. Note that we have used our modified definition of $\arh$ to obtain the above expression, due to which we have a $\gss{\trh}$ factor in the denominator instead of $\gss{0.204 \trh}$.

To obtain the value $\cs$ for a given EMDE case, we first obtain $x_f$ by solving Eq.~(\ref{focond}) with $a_f$ taken from Eq.~(\ref{af}). This $x_f$ value is then used in Eq.~(\ref{sigv_a}) to compute $\cs$. We find that this method yields $\cs$ that is within 10\% of the optimal $\cs$ value that solves the full Eqs.~(\ref{dmfo_eq}). We use this analytically obtained $\cs$ in our analyses in this work.

\section{Density Profiles for Cusps With $\rcore > \rcusp$}
\label{app:invcore}

\begin{figure}[h!]
\centering
\includegraphics[scale=1.4]{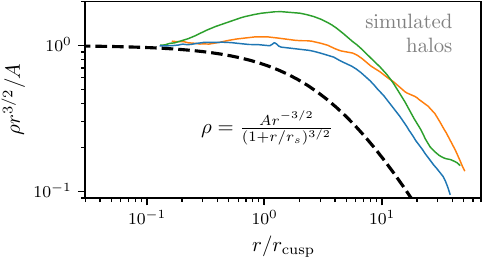}
\caption{Comparing the density profile in Eq.~(\ref{wi4}), with $r_s=e^{-3/2}\rcusp$, with those of the simulated halos H1 (orange), H2 (green), and H3 (blue) from Ref.~\cite{sten_survival}. The simulated halos are shown long after the initial collapse, at $a=6\ac$. These objects arise from a EMDE-like power spectrum with $\mathcal{P}\propto k^4$ up to the cutoff.}
\label{fig:sim-profiles}
\end{figure}

The core radius for a cusp is computed by assuming a $r^{-3/2}$ density profile and finding the radius at which the phase-space density of this profile exceeds $f_{\rm max}$, the maximum of the phase-space density set by DM freeze-out or gravitational heating. For some cusps, the core radius evaluated in this way can exceed the cusp radius which is set by the collapse time of the peak. To compute the core radii of such cusps, we need to assume a profile outside the cusp radius, which transitions from $r^{-3/2}$ behavior for small $r$ to a different power law for large $r$. We will consider the density profile \begin{equation} \label{wi4}
\rho_w (r) = \frac{Ar^{-3/2}}{\left(1 + r/r_s \right)^{3/2}},
\end{equation} where $r_s$ is the scale radius. This profile approaches the cusp profile $Ar^{-3/2}$ for $r \ll r_s$. We obtain $r_s$ by setting the integrated annihilation rate $J = \int \rho^2 \mathrm{d}V$ from $\rho_w$ equal to that from a cusp with profile $\rho(r) = Ar^{-3/2}$. Considering the volume integral from some radius $r_0$ to $\rcusp$ for the cusp profile, we have \begin{equation} \label{jcusp}
    J_{\rm cusp} = 4 \pi \int_{r_0}^{\rcusp} A^2 r^{-3} r^2 \mathrm{d} r = 4 \pi A^2 \ln (\rcusp / r_0),
\end{equation} while for our extended profile, we have \begin{equation}
    J_w = 4 \pi \int_{r_0}^{\infty} \rho_w^2(r) r^2 \mathrm{d}r = 4 \pi \rho_0^2 r_s^3 \left[ \ln \left( 1 + \frac{r_s}{r_0} \right) - \frac{r_s(3r_s + 2r_0)}{2 (r_s + r_0)^2}    \right].
\end{equation} Now, for $r_0 \ll r_s$, $J_w = 4\pi A^2 \ln (e^{-3/2} r_s/r_0)$. Equating this with the $J$ given by Eq.~(\ref{jcusp}) yields $r_s = \rcusp e^{-3/2}$.
Figure~\ref{fig:sim-profiles} shows how the resulting density profile compares to simulation results from Ref.~\cite{sten_survival}. It is evidently a very conservative choice, lying below the density profiles of the simulated halos at all radii.

To find the core radius for this profile, we look at the phase-space density associated with it. We assume an isotropic distribution function so that the phase-space density $f$ depends only on the energy $E$. For each radius, $f(E)$ is maximized at zero velocity, for which $E = \phi$, the gravitational potential.
The gravitational potential is\begin{equation}
    \phi_w(R) = 8\pi G A r_s^{1/2} \left\{ 1 - \left( \frac{1 + R}{R} \right)^{1/2} + \frac{1}{2R} \ln [1 + 2R + 2(R + R^2)^{1/2}] \right\}
\end{equation} with $R = r/r_s$, and we use the fitting function for $f(E)$ given by Ref.~\cite{widrow2000}. We invert $f(E = \phi_w) = \fmax$ to find the core radius, where $\fmax$ is set at DM freeze-out or by gravitational heating.

With the core radius calculated by inverting the above relation, we now assume the density profile \begin{equation}
    \rho(r) = \begin{cases}
    \rho_c = \rho_w(\rcore), & r < \rcore \\
    \rho_w(r). & r > \rcore
    \end{cases}
\end{equation} The annihilation J-factor, given by the integral of $\rho^2 \mathrm{d}V$, is then \begin{equation}
    \frac{J}{4 \pi} = \frac{\rho_c^2 \rcore^3}{3} + A^2\left[ \ln \left( \frac{r_s + \rcore}{\rcore} \right) - \frac{r_s(3r_s + 2\rcore)}{2(r_s + \rcore)^2} \right].
\end{equation} 

\section{Cusp survival in EMDE cosmologies}\label{sim-survival}

The prompt cusp distribution discussed in Sec.~\ref{cusps} is evaluated by associating a prompt cusp with each peak in the initial density field. However, some peaks do not collapse to make prompt cusps, because they are accreted onto another structure first.
Also, halos merge together over time, and it is possible that during this process, their central prompt cusps also merge.
Using the full-box (not zoomed) cosmological simulations of Ref.~\cite{sten_survival}, Ref.~\cite{sten_white_cusp} estimated that approximately half of the initial peaks can be associated with prompt cusps at late times.
These simulations do not resolve the cusps in detail. Instead, the estimate is based on associating initial peaks with bound clumps of matter identified in the simulations using the \textsc{subfind-hbt} algorithm implemented in the \textsc{Gadget-4} simulation code \cite{Springel:2020plp}.
If the particles comprising an initial peak can be associated with a unique bound object that is not already associated with a different peak, then the peak is determined to have collapsed to form a cusp located within that bound object.
We can then track the survival of the bound object as a proxy for the survival of the cusp.
However, bound subhalos are known to be artificially fragile in numerical simulations \cite{vandenBosch:2017ynq}, and recent numerical \cite{Errani:2019sey} and analytic \cite{Amorisco:2021hch,2022MNRAS.517.1398B,2022MNRAS.516..106D,Stucker:2022fbn,2024arXiv240203430R} studies indicate that a subhalo with a density cusp cannot be destroyed by tidal forces from a smooth host halo.
Therefore, a cusp should only be assumed to be destroyed if the bound clump associated with it disappears from the simulation while it is close to the central cusp of a host halo.

Reference~\cite{sten_survival} simulated cosmologies with linear matter power spectra of the form $P(k)\propto k^n\e^{-(k/k_\mathrm{cut})^2}$ for $-3<n\leq1$ and some cut-off wavenumber $k_\mathrm{cut}$, and Ref.~\cite{sten_white_cusp} analyzed cusp survival for the $n=-2$ and $n=-2.67$ cases, which resemble standard cold and warm dark matter cosmologies. However, Ref.~\cite{sten_survival} also simulated the $n=1$ case, which closely resembles the matter power spectra that arise from early matter domination (see Fig.~\ref{fig:psemde}).
In this appendix, we repeat the cusp survival analysis for the $n=1$ cosmology.
As in Ref.~\cite{sten_white_cusp}, we define the mass unit for this analysis to be the mass of a typical prompt cusp, $\tilde m\equiv 7.3\langle k^2\rangle^{-3/2}\bar\rho$, where $\bar\rho$ is the mean comoving matter density and $\langle k^2\rangle\equiv\int{\diff^3 \vec k}~k^2 P(k)/\int{\diff^3 \vec k}~P(k)=2k_\mathrm{cut}^2$ (for $n=1$).
The simulation particle mass and box size are 
$\tilde m/10.4$ 
and $1.03\times 10^8\tilde m$, 
respectively.

\begin{figure}[h!]
\centering
\includegraphics[width=\textwidth]{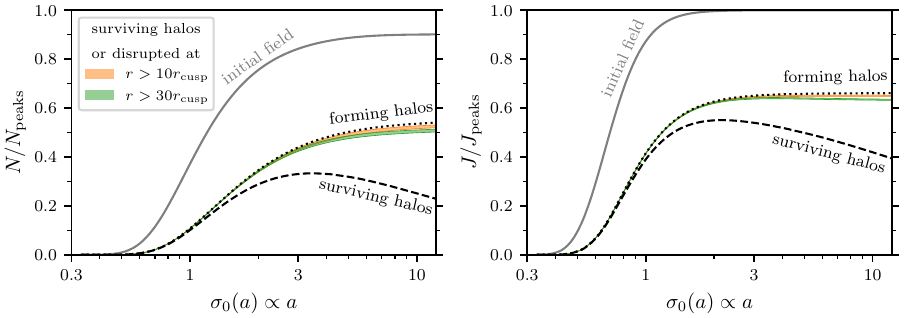}
\caption{Estimated fraction of initial peaks that can be associated with surviving prompt cusps, based on the simulation in Ref.~\cite{sten_survival} of an EMDE-like cosmology. The left-hand panel shows the normalized count of cusps, while the right-hand panel shows their normalized contribution to the annihilation rate. Both are shown as a function of the linear-theory rms density variance $\sigma_0$, which is a time parameter proportional to the scale factor $a$. In each panel, the solid curve shows the prediction based on the collapse times of all peaks in the initial density field. The dotted curve shows the contribution from peaks that are inferred to actually collapse because they can be associated with unique bound objects in the simulation. The dashed curve shows the contribution from peaks associated with surviving bound objects, but simulated subhalos are known to be artificially fragile, so the green and orange bands correct for this effect based on a range of assumptions, as described in the text of Appendix~\ref{sim-survival}.}
\label{fig:sim-survival}
\end{figure}

Figure~\ref{fig:sim-survival} shows the result of this analysis (c.f. Fig.~11 of Ref.~\cite{sten_white_cusp}; precise details of the analysis procedure can be found in Appendix B of Ref.~\cite{sten_white_cusp}).
The two panels show the number of cusps and the annihilation rate in cusps, respectively, both as a function of time.
The solid curve shows the prediction from the peaks in the initial density field, by accounting only for when they are predicted collapse.
The dotted curve shows the peaks that are interpreted to collapse into bound objects, by only counting those that are associated with unique \textsc{subfind-hbt} subhalos. A little over half of the peaks that are predicted to collapse actually do, by this estimate, and the cusps that result from those peaks contribute around two thirds of the total predicted annihilation rate.
Also, there is a significant time delay between the predicted (solid curve) and inferred (dotted curve) collapse times, but this is an artificial consequence of the simulation resolution.
\textsc{subfind-hbt} only identifies bound clumps of more than 20 particles, and the mass of a typical cusp corresponds to around 10 simulation particles. Thus, a cusp can only be identified by the algorithm after it grows a substantial halo around it. A consequence of this resolution limitation is that the fraction of collapsing peaks (dotted curve) is likely underestimated. For example, this analysis would miss a peak that collapsed into a cusp but failed to grow a significant halo around it before falling into another halo.

The dashed curves in Fig.~\ref{fig:sim-survival} indicate the contribution from initial peaks whose associated \textsc{subfind-hbt} bound objects persist in the simulation.
As noted above, the cusps of these objects can only physically be destroyed by sinking into the center of a host halo. Therefore, the green and orange bands mark the contribution from peaks whose \textsc{subfind-hbt} halos either survive in the simulation or vanish farther than some threshold distance from the center of a host halo. For the orange band, the distance threshold is 10 times the larger of the radii $r_\mathrm{cusp}$ of the subhalo and host prompt cusps, while for the green band, the distance threshold is 30 times that radius. Note that to account for the lower resolution of the $n=1$ simulation, we pick larger distance thresholds than Ref.~\cite{sten_white_cusp} did, meaning that we more readily allow subhalos to be destroyed. In principle, a subhalo's cusp should be destroyed only at a distance of order $r_\mathrm{cusp}$ from the center of the host. We pick much larger distance thresholds because, due to the low resolution of the simulation, the physical counterparts to artificially destroyed subhalos could still have sufficient mass to sink farther (due to dynamical friction) into the host halo.

It also sometimes occurs that a \textsc{subfind-hbt} object vanishes from the simulation when it is not a subhalo inside a host halo. There are two possibilities for characterizing these objects. There is no physical mechanism to destroy field halos, so the destruction can be assumed to arise from artificial evaporation due to gravitational collisions between simulation particles (e.g.~\cite{2003MNRAS.338...14P}). In that case, the cusp associated with the \textsc{subfind-hbt} object should be assumed to survive. Alternatively, the object could have been a collection of particles spuriously identified as a bound system, in which case the cusp should be assumed to have not formed in the first place. Each colored band in Fig.~\ref{fig:sim-survival} ranges between these two assumptions. Field halo disruption in the simulation is sufficiently rare that the different assumptions apparently have minimal effect.

Overall, the left-hand panel of Fig.~\ref{fig:sim-survival} indicates that around 55-60 percent of the peaks predicted to collapse can be associated with prompt cusps, according to this analysis. The right-hand panel indicates that these peaks yield around 65 percent of the total predicted annihilation signal, i.e., $f_s\simeq0.65$ in Eq.~(\ref{annihilation_per_mass}). We will continue to adopt the value $f_s \simeq 0.5$ suggested by Ref.~\cite{sten_white_cusp}, but this analysis shows that is a conservative choice for EMDE cosmologies.

\acknowledgments

We wish to thank Jens St\"ucker for the prompt responses and assistance with understanding his code, and Adrienne Erickcek and Dan Hooper for helpful discussions. H.G. is supported by NSF Grant AST-2108931. Computation for this work was done on the Hazel High Performance Computing Cluster at North Carolina State University.

\bibstyle{unsrt}
\bibliography{refs.bib}

\end{document}